\documentclass[journal,twoside,final]{IEEEtran}
\usepackage{graphicx}
\usepackage[cmex10]{amsmath}
\usepackage{amsmath}
\usepackage{amssymb, amsfonts}
\usepackage{algorithmic}
\usepackage{algorithm}
\usepackage{cite}

\usepackage{verbatim}

\ifCLASSOPTIONcompsoc
\usepackage[tight,normalsize,sf,SF]{subfigure}
\else
\usepackage[tight,footnotesize]{subfigure}
\fi

\usepackage{subfigure}
\graphicspath{{figures/}}

\begin{document}
\title{CoopGeo: A Beaconless Geographic Cross-Layer Protocol for Cooperative Wireless Ad Hoc Networks}

\author{\IEEEauthorblockN{Teck Aguilar, Syue-Ju Syue,~\IEEEmembership{Student Member,~IEEE}, Vincent Gauthier,~\IEEEmembership{Member,~IEEE}, Hossam Afifi, and Chin-Liang Wang,~\IEEEmembership{Senior Member,~IEEE}}

\thanks{
T. Aguilar, V. Gauthier, and H. Afifi are with the Department of Wireless Networks and Multimedia Services, Telecom SudParis, Evry, 91011, France (e-mail: \{teck.aguilar, vincent.gauthier, hossam.afifi\}@telecom-sudparis.eu).}
\thanks{
S.-J. Syue is with the Institute of Communications
Engineering, National Tsing Hua University, Hsinchu, Taiwan 30013,
Republic of China (e-mail: d949619@oz.nthu.edu.tw).}
\thanks{
C.-L. Wang is with the Department of Electrical Engineering
and the Institute of Communications Engineering, National Tsing
Hua University, Hsinchu, Taiwan 30013, Republic of China (e-mail: clwang@ee.nthu.edu.tw).}
\thanks{
This work was supported in part by the National Science Council of the Republic of China under Grants NSC 96-2219-E-007-008 and NSC 97-2221-E-007-005-MY3. This paper was presented in part at the 2010 IEEE 71st Vehicular Technology Conference (VTC 2010-Spring), Taipei, Taiwan, May 2010.}
}



\maketitle

\begin{abstract}
Cooperative relaying has been proposed as a promising transmission technique that effectively creates spatial diversity through the cooperation among spatially distributed nodes. However, to achieve efficient communications while gaining full benefits from cooperation, more interactions at higher protocol layers, particularly the MAC (Medium Access Control) and network layers, are vitally required. This is ignored in most existing articles that mainly focus on physical (PHY)-layer relaying techniques. In this paper, we propose a novel cross-layer framework involving two levels of joint design---a MAC-network cross-layer design for forwarder selection (or termed routing) and a MAC-PHY for relay selection---over symbol-wise varying channels. Based on location knowledge and contention processes, the proposed cross-layer protocol, CoopGeo, aims at providing an efficient, distributed approach to select next hops and optimal relays along a communication path. Simulation results demonstrate that CoopGeo not only operates properly with varying densities of nodes, but performs significantly better than the existing protocol BOSS in terms of packet error rate, transmission error probability, and saturated throughput.
\end{abstract}

\begin{IEEEkeywords}
Ad hoc networks, cooperative networks, cross-layer design, geographic routing, relay selection.
\end{IEEEkeywords}


\section{Introduction}
\IEEEPARstart{O}{ver} the last decade, there has been a tremendous wave of interest in the study of cooperative communications for wireless networks. By taking advantage of the broadcast nature of the wireless medium, neighbors overhearing data packets are allowed to assist in the ongoing transmission. Such resource sharing (e.g., power, antennas, etc.) among distributed nodes, which can increase the number of degrees of freedom (as introduced in \cite{Zheng2003}), is a fundamental idea of cooperative communications. Most attractively, without a centralized antenna array, cooperative systems are able to provide spatial diversity as well, in a distributed fashion.

Most existing work on cooperative techniques focuses on physical-layer cooperative relaying schemes, with various diversity-oriented signaling strategies proposed and demonstrated on the basis of information theory \cite{Nosratinia2004,Laneman2004,Laneman2003,Sendonaris2003,Kramer2005,SSL2007}. However, to achieve efficient communications while gaining full benefits from nodes cooperation, more interactions at higher layers of the protocol stack, in particular the MAC and network layers, are vitally required. Furthermore, an efficient cooperation-based MAC (or cooperative MAC) scheme should be not only payload-oriented but also channel-adaptive to improve the network throughput and diversity gain simultaneously; otherwise, an inefficient MAC scheme may even make cooperation gain disappear \cite{shan2009}.

Two major questions related to cooperative MAC design need to be answered: 1) when to cooperate? 2) whom to cooperate with and how to do selection? For the first question, intuitively cooperation may not be a requisite for reliable transmission if the direct link is of high quality. In addition, the use of cooperation inevitably introduces somewhat inefficiency due to extra protocol overhead and limited payload length.
Therefore a cooperative MAC protocol should be carefully designed to prevent unnecessary cooperation \cite{shan2009}.
In \cite{Ibrahim2008}, a cooperation metric related to the instantaneous source-relay and relay-destination channel measurements was proposed to decide if cooperation is needed.
The use of automatic repeat request (ARQ) and hybrid-ARQ schemes in cooperative networks has been discussed in \cite{Zhao2005, Dianati2006}.
In \cite{Laneman2004}, an incremental relaying protocol using limited feedback from the destination was proposed, which can be viewed as an extension of ARQ in the relaying context.
The second question about cooperative MAC design addresses the typical relay selection problem. There may exist a group of available relays around the source; however, some are beneficial and some not. How to find the optimal one(s) efficiently and effectively is of vital importance to a practical MAC protocol.
\IEEEpubidadjcol

Recent years have seen growing interest in the subject of relay selection \cite{shan2009,Ng2007,YK2008,Ibrahim2008,syue09,Liu2006,Beres2008,Bletsas2005,Gletsas2006,Guo2008, Liu2005,Zhu2006,Zhao2005,CTBTMA08,Gokturk2009,Yu2010,Wei2010}.
Some focus on the design of enhancing system reliability in a centralized manner \cite{Ng2007,YK2008,Ibrahim2008,syue09}, neglectful of the needs of overhead produced by nodes coordination as well as the feasibility of capturing lots of channel state information (CSI) among nodes.
To make relay selection more efficient, the authors of \cite{Liu2006} described how physical-layer cooperation can be integrated with the MAC layer to improve network performances. Other cross-layer issues are also included in \cite{Liu2006}.
In \cite{Bletsas2005,Gletsas2006,Guo2008}, distributed relay selection schemes based on the knowledge of local instantaneous channel conditions without requiring topology information are proposed.
CoopMAC \cite{Liu2005} and rDCF \cite{Zhu2006} are similar cooperative MAC protocols, which select a high-data-rate node to alleviate the throughput hindrance caused by low-data-rate nodes.
In \cite{Zhao2005}, a generalized concept of hybrid-ARQ is applied to relay networks, allowing that packet retransmissions could be performed at any appropriate relay.
In \cite{CTBTMA08}, the authors utilized busy tones to solve collisions in a cooperation scenario and to address the optimal relay selection problem.
In \cite{Yu2010}, the authors studied a fully opportunistic relay selection scheme under partial CSI for cellular networks, jointly considering macro and micro diversities.
In \cite{syue09}, we have proposed a geographic relay selection scheme based on the knowledge of location information of nodes. By jointly combining the source-relay and relay-destination distances, the optimal relay offering the best cooperative link can be efficiently determined. However, the selection process proposed by \cite{syue09} requires a central controller to decide which relay is most helpful, leading to more overhead and power consumption. One goal of this paper is to present a distributed relay selection protocol based on \cite{syue09}, with MAC-physical cross-layer design.

Likewise, in view of the interaction between the MAC and network layers, we also incorporate the routing issue into the design as a properly designed MAC protocol can facilitate routing process at the network layer, especially the beaconless geographic routing\footnote{\,Geographic routing can be applied to the Selection Diversity Forwarding \cite{Larsson2001,Souryal2005}---another way of achieving spatial diversity via forwarder selection---exploiting CSI to select routes with favorable channel conditions. However, we do not examine this channel-adaptive scheme in this paper. The diversity gain we discuss is only from relay selection in cooperative networks.} (BLGR) \cite{SRM2009,Heissen2003,Fubler2003,ZR2003,Blum2003,Casari2005,Sanchez2007}. BLGR is one of the most efficient and scalable routing solutions for wireless ad hoc and sensor networks. The key advantage of BLGR is that it needs neither prior knowledge of network topology for making a route decision nor the periodic exchange of control messages (i.e., beacons) for acquiring neighbors' geographic locations. A current node can make its own routing decisions by using local information.
In general, a BLGR protocol comprises two operating phases: forwarding phase and recovery phase. A forwarding node executes the greedy mechanism in the forwarding phase, and, if failing, switches to recovery mode to perform a face routing algorithm, finding another path to the destination.

It is noteworthy that BLGR at the network layer is usually coupled with MAC protocols to offer better network throughput and preserve advantageous properties such as localized operation and high scalability.
In \cite{Sanchez2007}, Sanchez \textit{et al} proposed a cross-layered BLGR protocol called BOSS, using a three-way (DATA/RESPONSE/SELECTION) handshake and an area-based timer-assignment function to reduce collisions among responses during the forwarder selection phase. Yet, as operating in the recovery mode, BOSS requires the exchange of complete neighborhood information for face routing. To avoid this drawback, we present a fully beaconless protocol without requiring beacons in both the greedy forwarding and recovery modes.

We have introduced above the roles of interactions between the MAC and physical layers and between the network and MAC layers in a cooperative scenario. In this paper\footnote{\,This paper is an extended version of our previous work \cite{AGSGAW2010}.}, we aim at investigating network-MAC-physical cross-layer design---with a focus on beaconless geographic protocols---to enhance overall system performance. Two issues, routing and relay selection, are the two chief considerations.
We assume that channels changes quickly enough as symbol-wise varying channels.
The proposed cross-layer framework, called CoopGeo, consists of two joint cross-layer designs: a joint network-MAC design for next hop selection and a joint MAC-physical design for relay selection. In particular, both the routing and relay selection solutions in CoopGeo are beaconless geographic protocols using contention-based selection processes, providing a strongly practical multi-layer integration for cooperative networks.

The contributions of this article are as follows:
\begin{itemize}
  \item We propose a distributed MAC-PHY cross-layer design for relay selection based on the centralized geographic approach in \cite{syue09}. 
  \item We present a fully beaconless approach to geographic routing with a MAC-network cross-layer design---both the greedy and recovery forwarding schemes need neither periodic exchange of beacons nor complete neighborhood information.
  \item The framework of CoopGeo, supporting localized operation as well as high scalability, is considerably practical for cooperative wireless ad hoc networks.
\end{itemize}

The rest of the paper is organized as follows. In Section II, we present the network model of cooperative networks along with the problem statement. Section III details the proposed CoopGeo with the cross-layer design for cooperative networks, in which beaconless geographic routing and relay selection, along with the protocol description, are included. In Section IV, we give some simulation results for CoopGeo and evaluate its performance by comparing with an existing protocol. Finally, we conclude this paper in Section V.

\section{Network Model and Problem Statement}
\subsection{Network Model} \label{Subsec:NetModel}
Consider a wireless ad hoc network of $k$ nodes randomly deployed in an area, expressed as a dynamic graph $G(V,E)$, where $V = \{v_{1},v_{2},\ldots,v_{k}\}$ is a finite set of nodes and $E = \{e_{1},e_{2},\ldots,e_{l}\}$ is a finite set of links between nodes. We denote a subset $N(v_i) \subset V$, $i = 1,\ldots,k$, as the neighborhood of the node $v_i$, i.e., those nodes within the radio range of $v_i$. In this paper, we consider there is a single session in the network, where data delivery may cross over multiple hops.

\begin{figure}[t!]
  \centering
  \subfigure[]{
    \label{fig:netmodel}
    \includegraphics[width=0.47\textwidth]{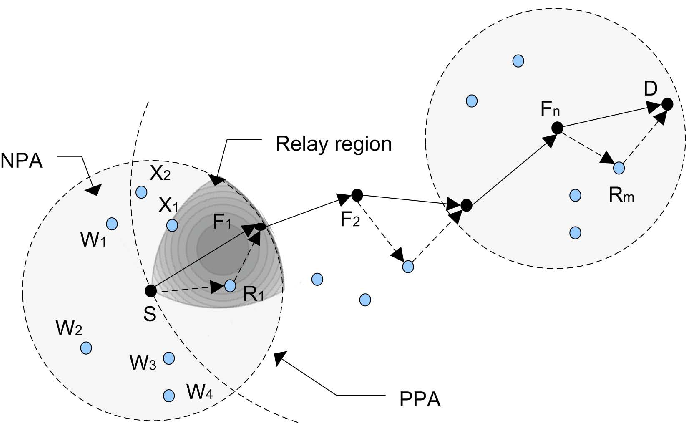}}     
  \subfigure[]{
    \label{fig:eachhop}
    \includegraphics[width=0.36\textwidth]{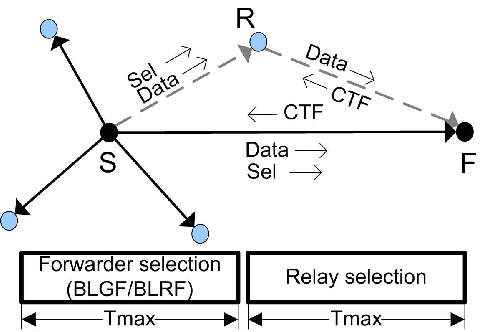}}      
  \caption{(a) Cooperative multihop ad hoc network model. (b) Direct and cooperative modes for each hop.}
  \label{fig:common_model}
\end{figure}

Fig. \ref{fig:netmodel} depicts the wireless ad hoc network model, in which the source $S$ sends its data to the destination $D$ in a multihop manner. In this figure the dashed circle centered at $S$ illustrates the radio range of $S$, and so on. At the beginning of every data transmission, $S$ broadcasts the data to its neighbors $N(S)$. One of these neighbors $N(S)$ is chosen as the next hop through a forwarder selection process, denoted as $F_1$. Two transmission modes, namely direct and cooperative modes, are considered to operate in each hop. In the direct mode, a point-to-point communication is performed by direct transmission; in the cooperative mode, it is done by cooperative relaying. The cooperative mode operates only when $F_1$ cannot correctly decode the data from $S$. After having a correct version of the data packet, $F_1$ acts as the source node and repeats the same procedure, and so on until the data packet reaches the destination $D$.

Since the multihop transmission is realized by concatenating multiple single-hop schemes as shown in Fig. \ref{fig:eachhop}, for convenience of notations we denote $S$ and $F$ as the current source and the forwarder (or called next hop), respectively. In addition, we represent $R_i$, $i=1,\ldots,|N(S)|$, as the candidate relays of $S$, one of which is going to cooperate with $S$ whenever needed.
In the following we introduce the signal models for the direct and cooperative transmission modes, respectively.

In the direct mode, $S$ broadcasts its symbol $x$ at the time index $i$ with transmission power $P$, where the average power of $x$ is normalized to unity. The received signals at $F$ can be expressed as
\begin{equation}
y_{S,F}^{(i)} = \sqrt{P} h_{S,F}^{(i)} x + n_{S,F}^{(i)},
\end{equation}
where $h_{S,F}^{(i)}$ is the channel coefficient from $S$ to $F$ and $n_{S,F}^{(i)}$ is the additive noise term. Throughout this paper, we assume that each node has a single antenna operating over frequency-flat Rayleigh fading channels and can only either transmit or receive data at any time slot. Moreover, the fading channels are assumed to be sufficiently fast-varying such that any channel coefficient, say $h_{u,v}^{(i)}$, modeled as $h_{u,v}^{(i)} \thicksim CN(0, \sigma_{u,v}^2)$, is constant over a symbol duration and may change from a symbol to another as an \textit{i.i.d.} random process. We also assume that all the channel coefficients among radio links are independent. Finally, we model all the noise terms as complex Gaussian random variables with zero mean and equal variance $N_0$, where, without loss of generality, we assume $N_0 = 1$.

For the cooperative mode, it applies a two-phase decode-and-forward (DF) strategy with single-relay selection, described as follows. In the first phase, $S$ broadcasts its symbol $x$ with transmission power $P_x$ while the next hop $F$ and a selected relay $R$ (through a relay selection process) listen. The received signals at $F$ and $R$ can be respectively expressed as
\begin{equation}
y_{S,F}^{(i)} = \sqrt{P_x} h_{S,F}^{(i)} x+ n_{S,F}^{(i)},
\end{equation}
\begin{equation}
y_{S,R}^{(i)} = \sqrt{P_x} h_{S,R}^{(i)} x + n_{S,R}^{(i)},
\end{equation}
where $h_{S,R}^{(i)}$ is the channel coefficient from $S$ to $R$ and $n_{S,R}^{(i)}$ is the additive noise term. In the second phase, with the simple adaptive DF strategy \cite{HZF2004}, the selected relay decides whether to forward the decoded symbol to the next hop. If the relay is able to decode the transmitted symbol correctly, it forwards the decoded symbol with identical transmission power $P_x$ to the next hop, and if not, it remains idle.
For practical use of this \textit{adaptive} mechanism, we consider that each relay is able to evaluate its own condition based on an SNR threshold. If the received SNR at the relay is greater than a certain threshold, the relay forwards; otherwise, it remains idle.
\begin{equation}
I_{R} = \left\{
\begin{array}{ll}
1, & \mbox{if } R \mbox{ decodes the symbol correctly}, \\
0, & \mbox{otherwise.}
\end{array} \right.
\end{equation}
Then, the received signals at the the next hop in the second phase can be written as
\begin{equation}
y_{R,F}^{(j)} = \sqrt{P_x I_{R}} h_{R,F}^{(j)} x + n_{R,F}^{(j)}, \, (j \neq i)
\end{equation}
where $h_{R,F}^{(j)}$ denotes the channel coefficient from $R$ to $F$ and $n_{R,F}^{(j)}$ denotes the AWGN.
Finally, the next hop coherently combines the received signals from the current source and the selected relay, i.e., $y_{S,F}^{(i)}$ and $y_{R,F}^{(j)}$, by using a maximum ratio combining (MRC)
\begin{equation}
y_F^{(j)} = \sqrt{P_x} h_{S,F}^{(i)*} y_{S,F}^{(i)} + \sqrt{P_x I_{R}} h_{R,F}^{(j)*} y_{R,F}^{(j)}.
\end{equation}
Consequently, the decoded symbol $\hat{x}$ at the next hop is given by
\begin{equation}
\hat{x} = \mbox{arg } \min_{x \in \mathcal{A}} \ |y_F - P_x (|h_{S,F}^{(i)}|^2 + I_{R}|h_{R,F}^{(j)}|^2) x|^2,
\end{equation}
where $|\mathcal{A}|=\Theta$ denotes the cardinality of $\Theta$-ary constellation.

By invoking the performance analysis in \cite{Su08}, the resulting symbol error rate (SER) at the next hop can be expressed as
\begin{equation}\label{Ps}
P_s \thickapprox \frac{4 N_0^2}{b^2 P_x^2 \sigma_{S,F}^2} \left( \frac{A^2}{\sigma_{S,R}^2} + \frac{B}{\sigma_{R,F}^2} \right),
\end{equation}
which is a tight approximation in a high SNR regime, where $b=\frac{3}{2(M-1)}$,
$A = \frac{M-1}{2M} + \frac{(1-1/ \sqrt{M})^2}{\pi}$, and $B = \frac{3(M-1)}{8M} + \frac{(1-1/ \sqrt{M)}^2}{\pi}$ in the case of $M$-QAM modulation.\footnote{\,The parameters $b$, $A$, and $B$ in the case of $M$-PSK modulation can be found in \cite{Su08} as well.}


Moreover, we make the following assumptions in the network model:
1) the network is dynamic and the network topology, including the cardinality of a node's neighborhood, the location of nodes, and the linkage between nodes, changes over time due to wireless environments, duty circles, and node failures, etc.; 2) each node is aware of its own location; 3) in addition to itself's location, the source knows the location of the destination, and so does any intermediate node; 4) all the network nodes are homogeneous, and each could become a source, relay, or forwarder.

\subsection{Problem Statement}\label{Statement}
In considering how cross-layer design improves network throughput and reliability for wireless cooperative ad hoc networks, the first question that arises concerns the joint MAC-network cross-layer routing design. For a network $G(V,E)$, given a source-destination pair $v_S, v_D \in V$, the objective of a routing task is to find a subset of forwarders $P_{F} = \{v_{F_1},v_{F_2},\ldots,v_{F_n}\}  \subset V$ that builds a routing path from $v_S$ to $v_D$ with successful packet delivery guaranteed. In particular, each forwarder in $P_F$ is determined locally, within a forwarding area defined as the radio coverage of the current source that is divided into a positive progress area (PPA) and a negative progress area (NPA), as shown in Fig. \ref{fig:areas}. The beaconless greedy forwarding (BLGF) and beaconless recovery forwarding (BLRF) are applied in the PPA and NPA areas, respectively.

The second question that this study addresses concerns the MAC-PHY cross-layer relay selection design. The aim of relay selection in CoopGeo is to find a subset of optimal relay nodes $P_{R} = \{v_{R_1},v_{R_2},\ldots,v_{R_m}\} \subset V\setminus P_F$ to enhance the network reliability, where each optimal relay $v_{R_i}$ that minimizes the average point-to-point SER for each cooperative hop is locally selected within a predefined relaying area.

One design goal of CoopGeo is to develop a fully beaconless approach to geographic routing that does not rely on periodic exchange of beacons as well as complete neighborhood information. Therefore, we consider that both the forwarder and relay selections use a locally-operated contention process based on location information and area-based timers. A specified interval of time $T_{max}$ is assigned to each selection process.
By tackling the above issues, we then contemplate a feasible cross-layer protocol that comprehensively integrates the network, MAC, and PHY layers to achieve a highly-efficient communication. In the following section we detail the framework of the proposed cross-layer design.

\section{CoopGeo: A geographic cross-layer protocol for cooperative wireless networks}
CoopGeo, in general, performs two tasks in wireless cooperative ad hoc networks: routing and relay selection. As described above, the routing process works in two phases, i.e. BLGF and BLRF. Both phases share equally a time interval $T_{max}$ within which the forwarder selection is executed. The first half of the $T_{max}$ period is allocated to the BLGF phase and the second half to the BLRF phase.

In the BLGF phase, a next hop that provides maximum progress toward the destination is selected through a timer-based contention process. As failing to find a next hop in the BLGF phase, the routing process enters transparently to the BLRF phase and applies face routing by using graph planarization along with a select-and-protest principle. Cooperative relaying is required after the routing task whenever the selected next hop decodes the data packet erroneously. In this case, CoopGeo starts out to execute the relay selection task within another time interval $T_{max}$, selecting an optimal relay that offers the best cooperative link between the current source and next hop.

Fig. \ref{fig:netmodel} gives an example for both the routing and relay selections in CoopGeo. The nodes competing in the BLGF phase are those located in PPA, i.e., $X_{1},X_{2},R_{1},$ and $F_{1}$. Those located in NPA, i.e., $W_{1},\ldots,W_{4}$, are considered to compete in the BLRF phase. The node $F_1$ is selected as the forwarder for the data transmission from the source $S$ to the forwarder $F_1$ that carries out a direct or cooperative transmission. In the case of cooperative transmission, the candidate relays with respect to the transmitter-receiver pair $(S,F_1)$ participate during the relay selection process are those within the relaying area (as will be defined later), including $R_1$ and $X_1$. As depicted in Fig. \ref{fig:netmodel}, $R_1$ is selected as the optimal relay node for the cooperative hop from $S$ to $F_1$.

\subsection{Beaconless Greedy Forwarding (BLGF)}\label{Greedy}
At the beginning of a data transmission, $S$ triggers the BLGF phase of the routing process by broadcasting its data to the neighborhood, while waiting for the best next hop's response during the first half of the $T_{max}$ time. During this period, the neighborhood compete to forward the message by setting their contention-based timers ($T_{CBF}$), as will be given in Section \ref{CBF}. When the best forwarder is selected due to its timer expiration, it sends a clear-to-forward (CTF) message to $S$, then the other candidates overhearing this message suppress their running timers and delete the data received from $S$. Since some candidates situated at the forwarding area may be unable to hear the CTF message, the hidden terminal problem could exist. To avoid this problem, $S$ broadcasts a confirmation message (SELECT) to indicate a forwarder's winning state; those hidden candidates overhearing it, will suppress their timers. As soon as receiving the SELECT message, the winning forwarder F sends an acknowledgement (ACK) to $S$ and, thereafter, it acts as the source and repeats the process hop-by-hop until the data is delivered to the final destination $D$.

\begin{figure}[t]
  \centering
  \includegraphics[width=0.45\textwidth]{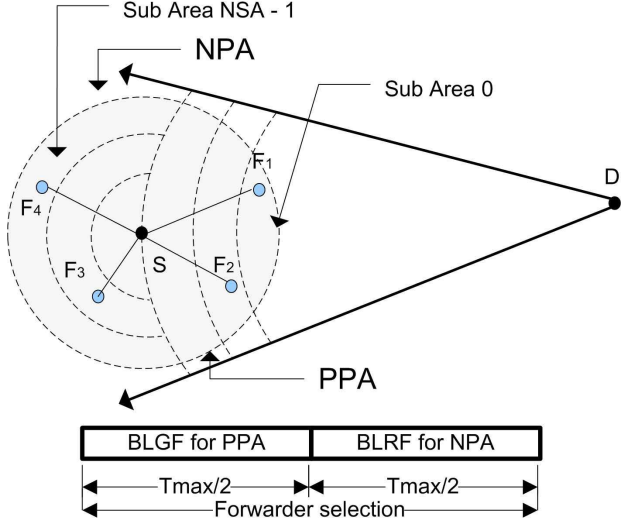}      
  \caption{Area division for CoopGeo routing. $F_{1}$ and $F_{2}$ are sub-area 0 and 1 of PPA respectively, whereas $F_{3}$ and $F_{4}$ are sub-area 4 and 5 of NPA respectively.
  }\label{fig:areas}
\end{figure}

\subsubsection{Geographic contention-based forwarder selection ($T_{CBF}$)}\label{CBF}
To carry out the BLGF mechanism, we base the timers settings on the metric proposed in \cite{Sanchez2007}, applying an area-based assignment function. Fig. \ref{fig:areas} depicts, as mentioned above, that the radio coverage of a current source is divided into the two areas PPA and NPA, both of which are further divided into sub-areas called Common Sub-Areas (CSAs) in order to avoid collisions during the contention period. Those candidate nodes situated at the same CSA offer similar progress toward $D$, and, hence, they have similar $T_{CBF}$ values. Note that unlike \cite{Sanchez2007}, we divide the NPA area by using concentric coronas instead of slides as used at the PPA area. We will discuss the reason in the BLRF section.

The timer setting for each candidate node is given as follows. First, each candidate node situated in PPA identifies which CSA group it belongs to by using the following equation:
\begin{equation}
CSA_{PPA} = \Big\lfloor NSA \times \frac{r - (d_{S,D} - d_{F_i,D})}{2r}\Big\rfloor \, ,
\end{equation}
where $NSA$ is a predefined even number of sub-areas to divide the coverage area, $r$ is the transmission range that is equal to the largest progress, and $(d_{S,D} - d_{F_i,D})$ represents the candidate's progress toward the destination.

Next, given $CSA_{PPA}$, hereafter called CSA, each candidate calculates its $T_{CBF}$ timer according to:
\begin{equation}\label{timer}
T_{CBF} = \Big(CSA \times \frac{T_{max}}{NSA}\Big) + rand \Big(\frac{T_{max}}{NSA}\Big) \, ,
\end{equation}
where $T_{max}$ represents the maximum delay time that the current source $S$ will wait for a next hop's response, and $rand(x)$ is a function of picking a random value between 0 and $x$ to reduce the collision probability. The $T_{CBF}$ function allocates the first half of $T_{max}$ to PPA candidates for the BLGF phase and the second half to the NPA candidates for the BLRF phase.

\subsection{Beaconless Recovery Forwarding (BLRF)}
As introduced before, the BLGF mode may suffer from the local minimum problem: the packet may be stuck at a node that does not have a neighbor (in PPA) closer to the destination than itself. To solve this problem, the Beaconless Forwarder Planarization (BFP) algorithm of \cite{Kalosha2008} that guarantees the packet delivery is applied at BLRF. BFP reduces the number of message exchanges by using the select-and-protest principle. In the select stage, some NPA neighbors are selected to form a planar subgraph according to a contention function, then, in the protest stage, false planar edges are removed from the subgraph. Finally, the traditional face routing algorithm is applied to select the forwarder.

BFP is performed in the BLRF phase of CoopGeo as follows. First, the current source detects the local minimum when a time interval of $T_{max}/2$ passes by without receiving any CTF message from neighbors in PPA. Thus, CoopGeo switches automatically from the BLGF mode to the BLRF mode, applying BFP during the second half of $T_{max}$. To accomplish this, the candidate nodes situated in the NPA determine their CSAs and compute their contention timers ($T_{CBF}$) for being used in the BFP algorithm. Once the planar subgraph is built, $S$ sends a SELECT message to the node that has been elected as a forwarder, which afterwards confirms the reception by sending back an ACK.

\begin{figure}[t]
  \centering
  \includegraphics[width=0.30\textwidth]{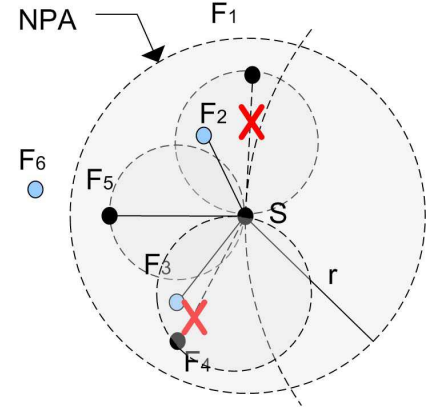}      
  \caption{Beaconless Recovery Forwarding in the area NPA as the Beaconless Greedy Forwarding fails.}\label{fig:RecoBFP}
\end{figure}

In \cite{Sanchez2007}, the CSAs of NPA are created according to the progress toward the destination. CoopGeo, by contrast, adopts the distance with respect to the node that is suffering the local minimum problem, and the slides are accordingly modified to concentric coronas. Thus, The NPA area is divided into $n = NSA/2$ equally-sized concentric coronas (as shown in Fig. \ref{fig:areas}), where the width of the $i$-th corona is $(\sqrt{i}-\sqrt{i-1})r_{1}$, and $r_{1}$ is the radius of the first corona calculated with $r_{1} = r / \sqrt{n}$. To use the same terminology as the one used in the BLGF phase, in the following a corona will be referred as a CSA. To set a contention timer, a candidate $F$ in NPA first finds its $CSA_{NPA}$ index by using the following equation:
\begin{equation}\label{CSAnpa}
CSA_{NPA} = \Big\lfloor \Big(\frac{\sqrt{n}\cdot d_{S,F}}{r}\Big)^2 \Big\rfloor + \frac{NSA}{2} \, .
\end{equation}
With the $CSA_{NPA}$ index, hereafter called CSA\footnote{\,A  $CSA$ value in the  forwarding selection is a nonnegative integer that falls in $[0,NSA - 1]$, where 0 corresponds to the area closest to $D$ and $(NSA -1)$ to the farthest one.}, each NPA forwarder candidate determines its contention timer according to \eqref{timer}, and then BFP is applied.

In this paper, we do not explain the BFP algorithm of \cite{Kalosha2008} in detail. Instead, we present an example to illustrate the procedures as in Fig. \ref{fig:RecoBFP}. Let us consider a scenario where the source $S$ is surrounded by six neighbors that respond in the order: $F_{1}$, $F_{4}$, and $F_{5}$ according to their timers defined by \eqref{timer}. $F_{2}$ receives the CTF message from $F_{1}$ and becomes a hidden node, $F_{3}$ receives the CTF from $F_{4}$, and $F_{6}$ receives the CTF from $F_{5}$. Thus, the hidden nodes are $F_{2}$, $F_{3}$ and $F_{6}$. $F_{2}$ is located in the proximity region (Gabriel Graph) of $F_{1}$ and $F_{3}$ in the proximity region of $F_{4}$. So, in the protest phase, $F_{2}$ protests against $F_{1}$ and $F_{3}$ protests $F_{4}$. Thus, $S$ removes the links with violating nodes (node in the proximity region of a node) and obtains a planar subgraph that will be used by the face routing algorithm to find the next forwarding node.

\subsection{MAC-PHY Cross-Layered Relay Selection}\label{CBR}
The relay selection process takes place after the forwarder selection whenever the demand for cooperation is announced by a forwarder. In this case, a new contention period will be allocated for relay selection.
The relay selection process, in this paper, is based on the selection criterion of \cite{syue09}, in which we had addressed a geographic relay selection problem.
Specifically, the best relay is selected according to a distance-dependent metric $m_i$, as shown in \eqref{STimer1}, relying on a combination of the source-relay and relay-destination distances.

Rewriting (\ref{Ps})---with a relay index $i$ introduced---in terms of coding gain and diversity order, we have
\begin{equation}
P_s = (\Delta \cdot \gamma)^{-d} \, ,
\end{equation}
where $\Delta$ denotes the coding gain of the scheme, given by
\begin{equation}\label{Delta}
\Delta = \sqrt{ \frac{b^2\sigma^2_{S,F}}{16} \left( \frac{A^2}{\sigma^2_{S,R_i}} + \frac{B}{\sigma^2_{R_i,F}} \right)^{-1} }	\, ,
\end{equation}
$d=2$ is the diversity order, and $\gamma = P/N_o$ represents the SNR, where $P=2 P_{x}$ is the total transmission power. 
The goal is to select the best relay that maximizes the coding gain $\Delta$ and, consequently, minimizes the SER. In \eqref{Delta}, the only term affecting the coding gain is
\begin{equation}\label{mi}
m_{i} \equiv \frac{A^2}{\sigma^2_{S,R_i}} + \frac{B}{\sigma^2_{R_i,F}}, \: i=1,2,...,N \, .  		
\end{equation}
Consider a $\sigma_{i,j}^2 \varpropto d_{i,j}^{-p}$ path loss model, where $p$ represents the path loss exponent. Then the channel variances $\sigma^2_{i,j}$ in (\ref{mi}) can be replaced with the distance-dependent parameters $d^{-p}_{i,j}$. Thus, (\ref{mi}) becomes
\begin{equation}\label{STimer1}
m_{i} = A^2 d^{p}_{S,R_i} + B d^p_{R_i,F}, \: i=1,2,...,N \, ,
\end{equation}
where $m_{i}$ is treated as our relay selection metric, which indicates the SER performance at the forwarder---the smaller the metrics is, the better the resulting SER performance will be. Therefore, the best relay can be determined according to the following criteria\footnote{\,Eq. \eqref{Ps} is a bound given as an asymptotically tight approximation at high SNR. As the SNR is sufficiently high, the average SER as in \eqref{Ps} is the same with the exact SER. For low SNRs, although \eqref{Ps} does not hold anymore, it does not affect the correctness of the selection for the best relay (or the second-best, third-best relays, and so on).}:
\begin{equation}\label{RScriterion}
i^* = \mbox{arg }\min_{i \in \{1,..,N\}}m_i =  \mbox{arg }\min_{i \in \{1,..,N\}} A^2 d_{S,R_i}^p + B d_{R_i,F}^p \, .
\end{equation}

We note that the best relay selected by the above criterion is the one that provides the best source-relay-forwarder cooperative link in terms of average SER at $F$. The relay selection process in \cite{syue09}, however, requires a central controller to make a best-relay decision according to the responses from all candidate relays. To reduce the required overhead while achieving a more efficient relay selection process, we propose a distributed relay selection protocol using MAC-PHY cross-layer design, as presented in the following part.

\subsubsection{Geographic contention-based relay selection}
The selection process starts as soon as each candidate relay overhears the DATA/CTF packets. Each candidate relay makes use of two relative distances $d_{S,R_i}$ and $d_{R_i,F}$ to calculate its own selection metric according to \eqref{STimer1}. Here the path loss exponent is assumed as a known parameter. For the purpose of decentralization, the relay selection metric $m_i$ is encoded in time difference inside a timer-based election scheme. Once a candidate whose timer expiration occurs first, it relays the data packet to $F$, while the others candidates cancel their timers after receiving the packet. This contention-based relay selection scheme provides a distributed and efficient way to determine the best relay for each cooperative hop, answering one of the major questions about cooperative MAC design, i.e., whom to cooperate with and how to do selection? The metric defined in \eqref{STimer1} indicates the cooperative link quality in terms of average point-to-point SER, depending on the modulation types and the locations of nodes. In order to translate our relay selection metric \eqref{STimer1} into a timer, we normalize it according to relative distances from $\mathbf{x}^*$, which denotes the best placement of a relay (minimized the average point-to-point SER). We denote $\mathbf{x}_S, \mathbf{x}_F$, and $\mathbf{x}_i$ as the locations of the current source, the forwarder, and the $i$-th candidate relay, respectively. In addition, we define a mapping function $f$ that maps a candidate relay's location into its relay selection metric ($\mathbf{x}_S$ and $\mathbf{x}_F$ are fixed), as written in \eqref{STimer2}. The optimal point $\mathbf{x}^*$ can be obtained by solving the optimization problem \eqref{STimer3}. As a result, the best relay is the one whose metric is closest to $f(\mathbf{x}^*)$.
\begin{equation}
f(\mathbf{x}_i)  =  A^2 \left\| \mathbf{x}_i - \mathbf{x}_S \right\|^p + B \left\|\mathbf{x}_i-\mathbf{x}_F \right\|^p
\label{STimer2}
\end{equation}
\begin{equation}
\text{minimize} \quad f(\mathbf{x})  =  A^2 \left\|\mathbf{x}-\mathbf{x}_S \right\|^p + B\left\|\mathbf{x}-\mathbf{x}_F \right\|^p
\label{STimer3}
\end{equation}
\begin{equation}
\mathbf{x}^*  =  \frac{A^2 \mathbf{x}_S + B \mathbf{x}_F}{A^2+B}  ~~(\mbox{as } p=2) \, .
\label{STimer4}
\end{equation}

We then derive a mapping function $\mathcal{M}$, which scales our metric function $f$ into the interval $[0,1]$:
\begin{equation}
\mathcal{M}(f(\mathbf{x}))= \frac{ f(\mathbf{x}) - f(\mathbf{x}^*) }{ f(\mathbf{x}_{\max }) - f(\mathbf{x}^*)} \, ,
\label{STimer5}
\end{equation}
where $\mathbf{x}_{max}$ is defined as a point within the relaying area that is farthest away from $\mathbf{x}^*$.\footnote{\,$\mathbf{x}_{max}$ is used for normalization purpose. In (\ref{STimer5}), $f(\mathbf{x}_{\max })$ is theoretically the maximum relay selection metric. Fig. \ref{fig:STimer1a}, as an example, shows that the maximum value for $\mathbf {x}_{max}$ is the intersection between the transmission radius of $S$ and $F$.}
Finally, a contention timer at each candidate relay is set by using the following function:
\begin{equation}\label{Timer}
T_{CBR} = T_{max} \ \mathcal{M}(f(\mathbf{x}))  + rand \Big( \frac{2T_{max}}{NSA} \Big) \, .
\end{equation}

\subsubsection{Relay selection area}
\begin{figure}[bt]
\centerline{
\subfigure[]{
	\includegraphics[width=0.23\textwidth]{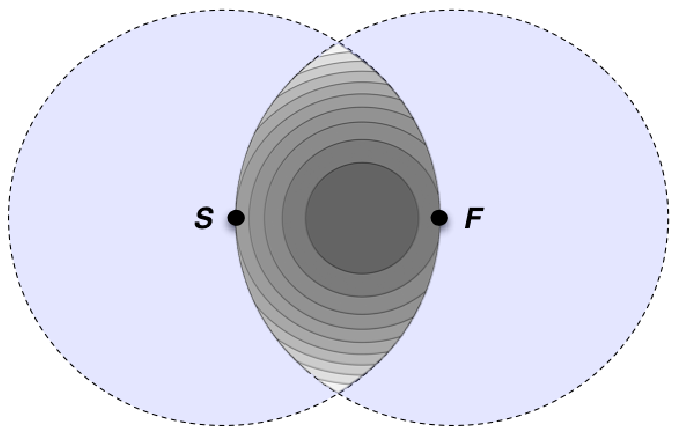}      
	\label{fig:STimer1a}
}
\hfil	
\subfigure[]{	
 	\includegraphics[width=0.23\textwidth]{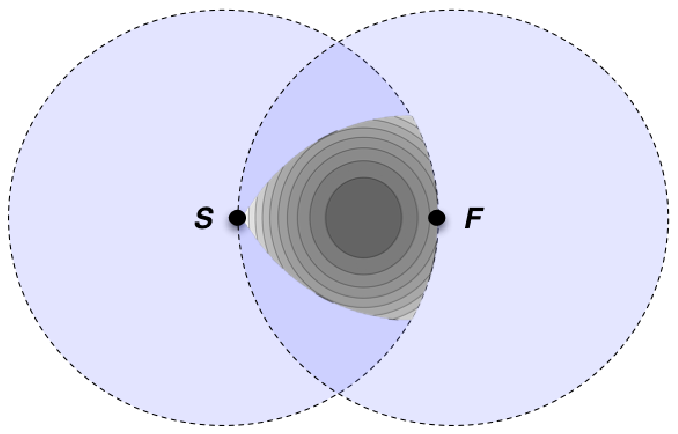}      
	\label{fig:STimer1b}
}}
\caption{Mapping of the relay selection metric onto (a) the set $\mathcal{C}$ and (b) the set $\mathcal{D}$ for a normalized distance between the current source $(0,0)$ and its next hop $(1,0)$.}
\end{figure}

The CoopGeo relay selection process do not use control messages as in the forwarding selection process so as to guarantee that only one node has been selected as relay, thus avoiding message duplications or collisions.
Once overhearing the relayed message from the selected relay, other candidate relays suppress their contention timers. Since the candidates should be located within a predefined area for the execution of relay selection, we consider the relaying area size as a way to control the corresponding impact. The relaying area is formed by the positions of the current source and forwarder.
Fig. \ref{fig:STimer1a} and \ref{fig:STimer1b} illustrate two relaying areas.
Firstly, let the set $\mathcal{C}$ represent a relaying area formed by the intersection of the current source and forwarder's coverage areas. Secondly, we denote the set $\mathcal{D}$ as another relaying area shaped in a Reuleaux triangular form, from the source's point of view.
In the first case, for any candidate relay $\mathbf{x}_{i} \in \mathcal{C}$, its selection metric is mapped onto this set, with $\mathcal{M}(f(\mathbf{x}_{i})) \in [0,1]$. For the Reuleaux triangle, any candidate $\mathbf{x}_{i}$ and any other possible one $\mathbf{x}_{j}$ have the following relationship: $\left\| \mathbf{x}_{i} - \mathbf{x}_{j} \right\|^{2} \leq r , \forall \mathbf{x}_{i}, \mathbf{x}_{j} \in \mathcal{D}, i \neq j$, where $r$ is the transmission range of a node. Hence, from the relaying areas depicted in the figure, the Reuleaux triangular area is the best suited to be used since all candidate relays can hear each other, which, as a consequence, effectively avoids the hidden relay problem. It is obviously not this case for the intersection relaying area as in Fig. \ref{fig:STimer1a}.

\subsection{CoopGeo in Action}
\begin{figure}[t]
  \centering
  \includegraphics[width=0.47\textwidth]{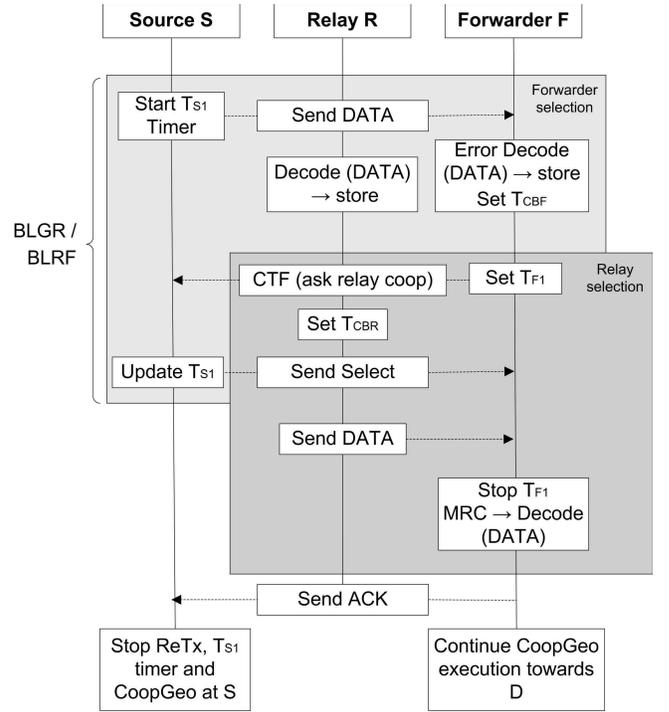}      
  \caption{CoopGeo in action.}\label{fig:inAction}
\end{figure}

In this subsection, we present the behavior of network nodes running CoopGeo, as depicted in Fig. \ref{fig:inAction}, for the data delivery from a source to a destination.
When the source $S$ intends to transmit its data to the destination $D$, it checks if the channel is free for a predefined time interval. If any, $S$ broadcasts its data packet DATA and starts a $T_{S1}$ timer. The neighbors of the source then receive the packet, store it, and set up their $T_{CBF}$ timers, as defined in \eqref{timer}, to participate in the forwarder selection process.\footnote{\,In geographic  protocols, the source generally has to indicate the location information of both itself and the destination in the packet header. The header added in the beginning of a packet is usually transmitted through low rate codes so that one could neglect its transmission error within the transmission range.}

The neighbor $F \in F_{i}$ whose timer expires first sends a CTF control message to claim the forwarding status, then it initializes a $T_{F1}$ timer. The other candidates overhearing this control message quit the forwarding selection process.
Here, the DATA/CTF handshake carried out by $S$ and $F$ is used to initiate the relay cooperation on demand. Specifically, $F$ indicates, in the CTF message, if relay cooperation is needed in case of error decoding. In this way, the neighbors situated in the relaying area formed by S and F and being able to correctly decode the DATA\footnote{\,Neighbors are supposed to determine the correctness of the DATA based on the measurements of received SNRs, as described in Sec. \ref{Subsec:NetModel}.}  start their $T_{CBR}$ timers, as defined in \eqref{Timer}, to participate in the relay selection process.

Upon receiving the CTF message, $S$ replies a SELECT message to $F$ for the confirmation of the forwarding status, while updating its $T_{S1}$ timer to the maximum allowed delay time for receiving an ACK from $F$. Meanwhile, the candidate relays start their $T_{CBR}$ timers. When the candidate $R \in R_{i}$ expires its $T_{CBR}$ timer in the first time, it becomes the relay node and immediately relays the stored data. The other candidates overhearing the data transmission stop competing in the relay selection process.
Finally, the forwarder combines the received signals from $S$ and $R$ with a maximum ratio combining, decodes the data, and stops its $T_{F1}$ timer. Then, it sends an ACK to S and continues the execution of CoopGeo toward $D$.

In addition to $T_{CBF}$ and $T_{CBR}$ timers, two more timers are used: $T_{S1}$ at the source and $T_{F1}$ at the forwarder. The timer $T_{S1}$ represents the maximum allowed time to find a forwarder toward D, given by
\begin{equation}
T_{S1} = T_{DATA} + T_{CTF} + T_{max} \, ,
\end{equation}
where $T_{DATA}$ and $T_{CTF}$ represent the data and CTF packet transmission times respectively and $T_{max}$ denotes the maximum time interval allowed for the forwarder selection process. For simplicity, in this equation we do not express the propagation delay.

Next, the timer $T_{S1}$ is updated for receiving an ACK from $F$. The setting of the updated $T_{S1}$, as given below, depends on whether relay cooperation is needed.
\begin{equation}
T_{S1} = \left\{
\begin{array}{lr}
T_{SEL} + T_{ACK}, \text{ if cooperation is not needed;}\\
T_{SEL} + T_{max} + T_{DATA} + T_{ACK}, \text{ otherwise}  ,
\end{array} \right.
\end{equation}
where the first statement includes the transmission time of the SELECT (from $S$ to $F$) and ACK (from $F$ to $S$) messages; the second statement includes the required time of the first statement plus the times $T_{max}$ and $T_{DATA}$ that correspond to the maximum allowed time for the relay selection and the time for relaying the packet, respectively.

As for $T_{F1}$, the affected value depends on whether the forwarder $F$ correctly decodes the received data from $S$, or whether relay cooperation is needed. For the former, $F$ listens to the channel and waits for a SELECT message from $S$, which completes the direct communication mode; for the latter, $F$ waits for the SELECT message and relayed DATA from the source and the relay, respectively. The timer setting of $T_{F1}$ is expressed as follows:
\begin{equation}\label{decode}
T_{F1} = \left\{
\begin{array}{lr}
T_{CTF} + T_{SEL} ,\text{ if cooperation is not needed;}\\
T_{CTF} + T_{SEL} + T_{max} + T_{DATA}, \text{ otherwise}  ,
\end{array} \right.
\end{equation}
where the first statement allocates the time required to transmit the CTF message as well as the time required to receive a SELECT message from $S$; the second includes the time of the first statement, plus the maximum allowed time for relay selection and the time for relaying the data. Also, $T_{F1}$ does not consider the propagation delay.

If the timer $T_{S1}$ of $S$ expires before receiving a CTF or an ACK from $F$, there are different possibilities: 1) $S$ could not find a forwarder; 2) $F$ could not receive the SELECT message from $S$; 3) $F$ could fail; 4) $F$ could not receive the data packet from $R$ in the cooperative mode. For these situations, the CoopGeo protocol is restarted.
In addition, it can be seen that the two most significant timers are $T_{CBF}$ and $T_{CBR}$, which are used to select a forwarder $F$ and an optimal relay $R$ in each hop through contention mechanisms; the use of the timers $T_{S1}$ and $T_{F1}$ are to help detect a problem during the CoopGeo execution.

\begin{figure}[t]
  \centering
  \includegraphics[scale=0.68]{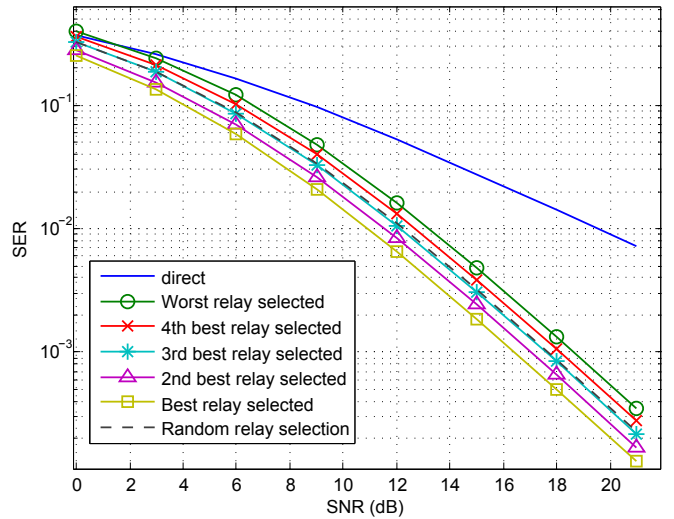}
  \caption{Performance comparison for relay selection when using from the best to worst relays. }\label{fig:relaysel}
\end{figure}

\begin{table}[t]
\renewcommand{\arraystretch}{1.3}
\caption{Simulation Settings}
\label{table2}
\centering
\begin{tabular}{c|c|c|c}
\hline
\bfseries Input & \bfseries Value  & \bfseries Input & \bfseries Value \\
\hline\hline
No. of Neighbors & 1-20 & Tx. Power & 25 dBm \\
Channel Model & Rayleigh & Average Noise & 20 dB \\
Pass Loss Exp. & 2 & Noise Figure & 15 dB \\
Carrier Freq. &  2.412 GHz & Packet Size & 1538 Bytes \\
Channel BW & 22 MHz & Contention Period & 500 $\mu s$ \\
Modulation Type & QAM & No. of Topologies & 20000 \\
Constellation Size  & 4-128 & No. of Trials & 2000000 \\
\hline
\end{tabular}
\end{table}

\begin{figure*}[!ht]
\centering
\subfigure[]{
\label{fig:ex3-a}%
\includegraphics[width=0.48\textwidth]{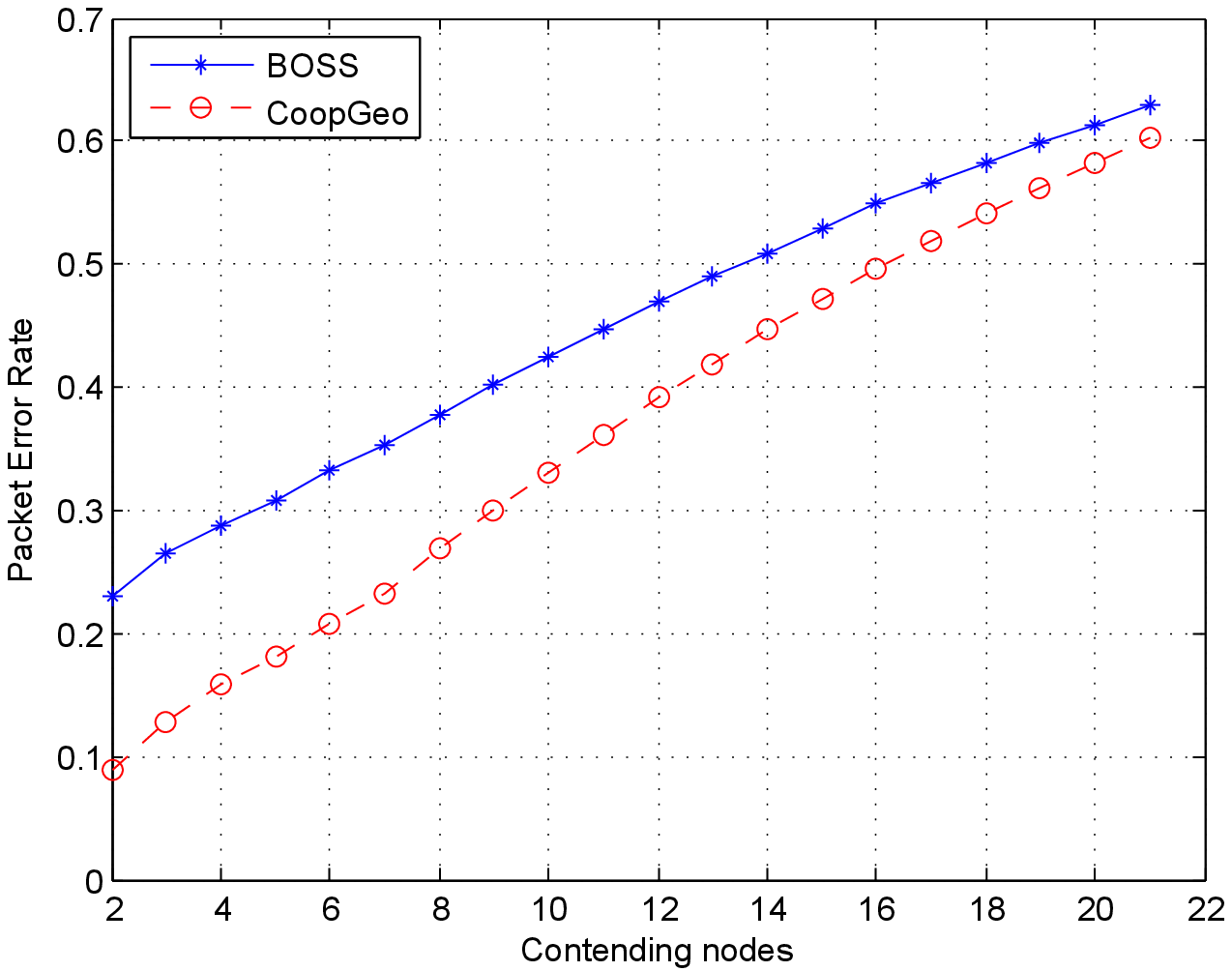}}
\hspace{8pt}%
\subfigure[]{
\label{fig:ex3-b}%
\includegraphics[width=0.48\textwidth]{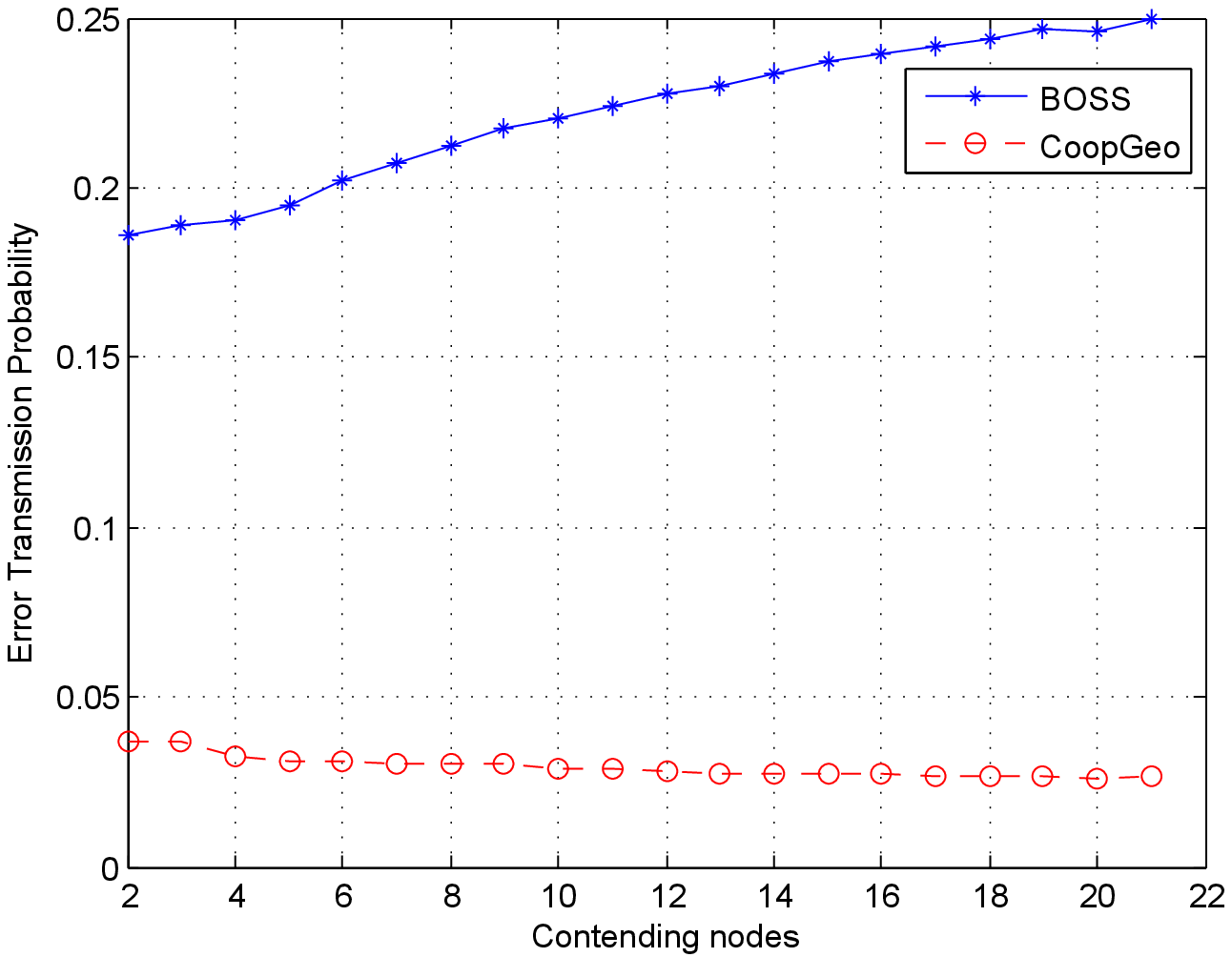}}\\      
\hspace{8pt}%
\subfigure[]{
\label{fig:ex3-c}%
\includegraphics[width=0.47\textwidth]{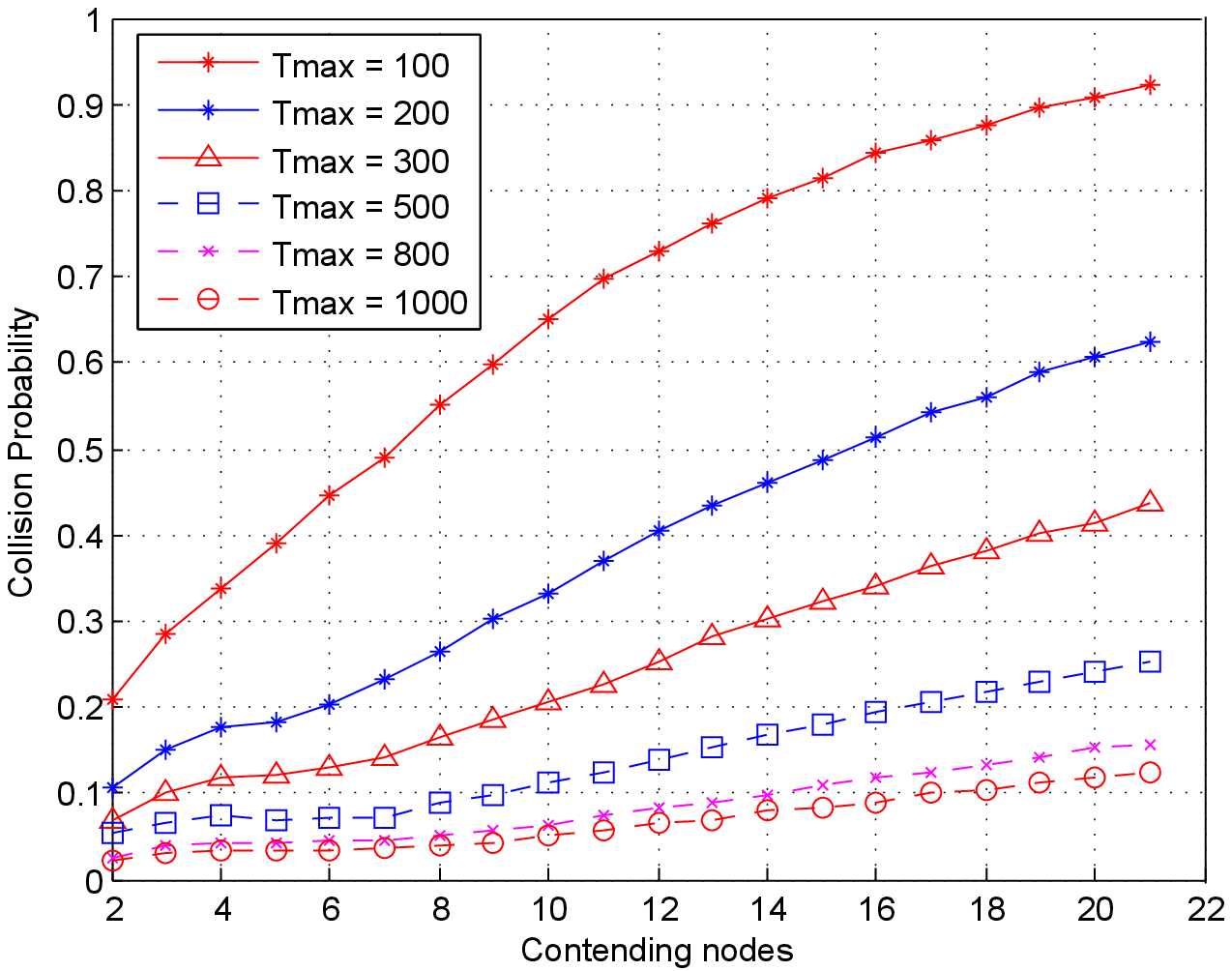}}      
\hspace{8pt}%
\subfigure[]{
\label{fig:ex3-d}%
\includegraphics[width=0.48\textwidth]{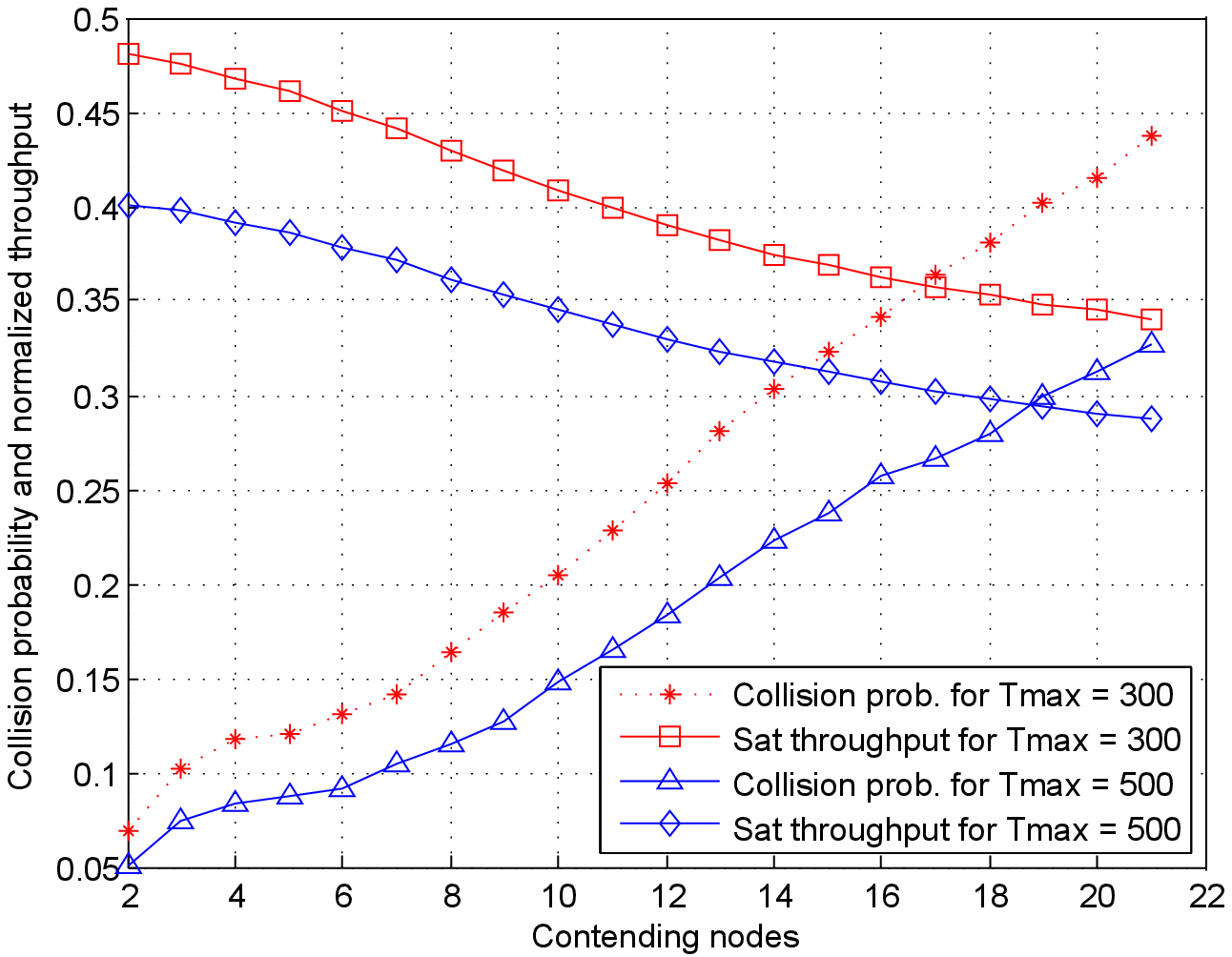}}\\      
\hspace{8pt}%
\subfigure[]{
\label{fig:ex3-e}%
\includegraphics[width=0.51\textwidth]{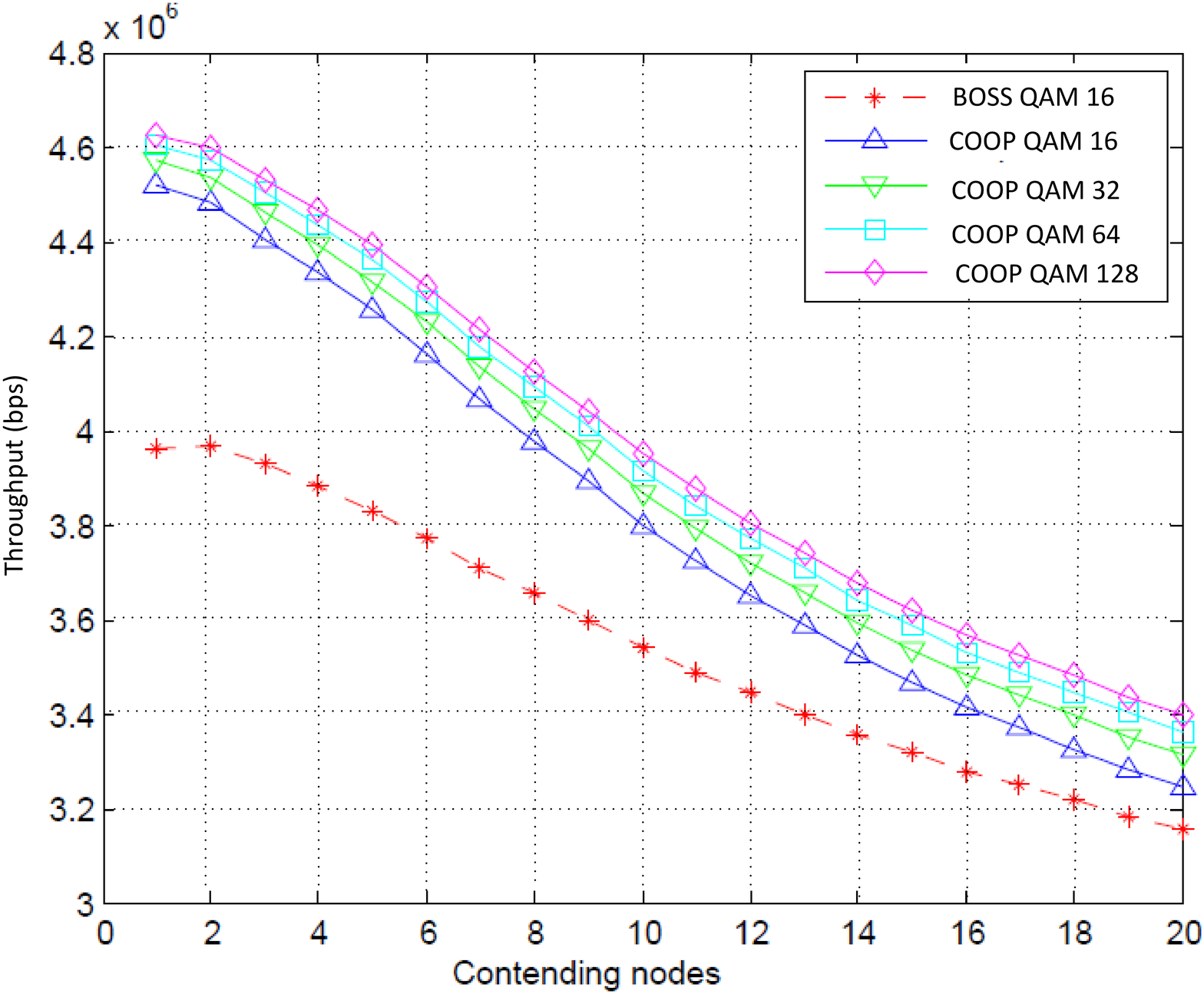}}\\      
\caption{
\subref{fig:ex3-a} Packet error rate for $T_{max}$ = 500$\mu s$.
\subref{fig:ex3-b} End-to-end transmission error probability for $T_{max}$ = 500$\mu s$.
\subref{fig:ex3-c} CTF-Relayed message collision probability with changing $T_{max}$ from 100$\mu s$ to 1000$\mu s$.
\subref{fig:ex3-d} Normalized saturated throughput and collision probability for $T_{max}$ = 300$\mu s$ and $T_{max}$ = 500$\mu s$. 
\subref{fig:ex3-e} CoopGeo saturated throughput for different QAM modulation types.}
\label{fig:ex3}%
\end{figure*}

\section{Performance Evaluation}
We first consider a single-hop cooperative relay network with $N=5$ available relays, deployed in $\mathbf{R}^2$. Denote $(x,y)$ as the coordinates of nodes. We locate the source and the destination at $(0,0)$ and $(1,0)$ respectively, and randomly place, with uniform distribution, the relays in a square field following $\{(x,y) \, | \, 0 \leq x \leq 1, |y| \leq 0.5\}$. We assume that the channel variances between any two nodes follow $\sigma_{i,j}^2 \varpropto d_{i,j}^{-p}$, where the path loss exponent is taken to be $p=2$ in our simulations. The channel variance is normalized to unity for unit distance. QPSK modulation is used in this simulation and the fading channels are assumed sufficiently fast-varying such that the channel coefficients keep constant only within every symbol interval. The number of network topologies is 200.

For each realization of nodes distribution, we can determine the distances from each relay to the source as well as destination, and then the corresponding selection metric for each relay can be determined using \eqref{RScriterion}. According to the selection criterion as introduced in Sec. \ref{CBR}, the best relay is the one with the minimum selection metric, while the second best relay has the second minimum selection metric and so on.

Fig. \ref{fig:relaysel} depicts the SER versus SNR performance of the above scenario, where SNR is defined as $P/N_0$ and $P$ is the total transmit power fixed. In Fig. \ref{fig:relaysel}, the performance of direct transmission from the source to the destination is provided as a benchmark for a non-cooperation scheme. Fig. \ref{fig:relaysel} shows that the selected best relay contributes to the minimum SER at the destination as compared to other relays. In addition, it also reveals that worse relays corresponds to larger selection metrics, that is, the smaller the selection metrics, the better the resulting SER performances. Thus, we have demonstrated that by using the geographical information, nodes in cooperative networks can efficiently perform relay selection to improve the SER performance at the destination. Moreover, we also compare the performance with a possible relay selection approach, named random relay selection, which means that the source randomly selects a cooperating relay without any information for each transmission. We see, in Fig. \ref{fig:relaysel}, that the performance curve of the random selection scheme lies in between the best and worst selections. This is because each relay has the same opportunities to be selected such that the performance will be averaged over all the distributed relays.

The next step in our simulation methodology is to evaluate the PHY/MAC layer performance of CoopGeo with Monte-Carlo simulations. We simulated the three lower layer processes, and our simulation settings are given in Table \ref{table2}. Our results are based on 20,000 randomly generated topologies, where all the nodes are competing to access the channels. We start by solving the two problems as stated in Section \ref{Statement}. Once the forwarder and relay sets are obtained, we use them to evaluate the packet error rate, average error transmission probability, saturated throughput, and some others with varying the input parameters.

\subsection{Packet Error Rate and Transmission Error Probability}
In Fig. \ref{fig:ex3-a}, we show the average packet error rate performance for two different protocols: CoopGeo and BOSS\cite{Sanchez2007}. The packet error rate includes both the probabilities of collision within different contention periods and transmission error over wireless channels. Fig. \ref{fig:ex3-a} shows that CoopGeo achieves a lower packet error rate 2.5 times less than the traditional geographic protocol BOSS in the best condition. Also, we see that the packet error rate performance curves of the two get closer with increasing the number of nodes in the neighborhood.
Moreover, in Fig. \ref{fig:ex3-b}, we show that CoopGeo improves significantly the average error transmission probability with increasing the number of contending nodes. This is due to the accurate selection of the relay when more nodes are present in the neighborhood. It is noted that the CoopGeo experiment offers a very low transmission error rate, which can be used to raise the constellation size of modulation to improve the bandwidth efficiency without loss of end-to-end throughput.

\subsection{Results with Varying Input Parameters}
\subsubsection{Varying the contention window $T_{max}$}
In this simulation, we assess the impact of the contention window size $T_{max}$ (that controls the delay time of a contending node when it tries to forward/relay a packet) on the CoopGeo performance. we first simulate our protocol with different values of $T_{max}$ from $100\mu s$ to $1000\mu s$. In Fig. \ref{fig:ex3-c}, we see that the collisions, caused by the contending nodes when they send their CTF packets or relayed messages, reduce with increasing the $T_{max}$ size.  The sizes varying from $500\mu s$ to $1000\mu s$ are the best suited for CoopGeo, as they achieve much lower collision probability as compared with the other cases. From the relationship between the normalized throughput with cooperative communications and the CTF-relayed messages collision probability, we observe that we may use a smaller $T_{max}$ size without affecting the performance of the protocol when fewer contending nodes are used for the case of $T_{max} = 300\mu s$, as shown in Fig. \ref{fig:ex3-d}. By taking $T_{max} = 500\mu s$ from the previous result as a reference, it can be seen that for a smaller saturated throughput rate with respect to $T_{max} = 300\mu s$, we may handle scenarios with higher densities.

\subsubsection{Varying the constellation size}
Finally, in Fig. \ref{fig:ex3-e}, we provide the saturated throughput of CoopGeo and compare it with the traditional geographic MAC-routing approach BOSS (that uses direct transmission). Fig. \ref{fig:ex3-e} shows that the CoopGeo outperforms the traditional scheme in terms of saturated throughput, with different constellation sizes used. Due to very low transmission error rate in the cooperation-based CoopGeo scheme, one can increase the constellation size according to different transmission environments without deteriorating the end-to-end throughput.

\section{Conclusions}
In this paper, we have proposed a cross-layer protocol CoopGeo based on geographic information to effectively integrate the network, MAC, and PHY layers for cooperative wireless ad hoc networks. The CoopGeo provides a MAC-network cross-layer protocol for forwarder selection as well as a MAC-PHY cross-layer protocol for relay selection. Both the selection schemes are based on location information of the nodes without periodic exchange of beacons and complete neighborhood information. Simulation results demonstrate that the proposed CoopGeo operates well with different densities and achieves better network performances than the existing protocol BOSS in terms of packet error rate, transmission error probability, and saturated throughput. Due to the beaconless local operation property, the CoopGeo is highly efficient and scalable to any changes in the network topology.
\section*{Acknowledgment}
The authors would like to thank the associate editor and the anonymous reviewers for their comments and suggestions, which have led to an improvement of this paper.

\bibliography{bare_jrnl}

\begin{thebibliography}{10}
\providecommand{\url}[1]{#1}
\csname url@samestyle\endcsname
\providecommand{\newblock}{\relax}
\providecommand{\bibinfo}[2]{#2}
\providecommand{\BIBentrySTDinterwordspacing}{\spaceskip=0pt\relax}
\providecommand{\BIBentryALTinterwordstretchfactor}{4}
\providecommand{\BIBentryALTinterwordspacing}{\spaceskip=\fontdimen2\font plus
\BIBentryALTinterwordstretchfactor\fontdimen3\font minus
  \fontdimen4\font\relax}
\providecommand{\BIBforeignlanguage}[2]{{%
\expandafter\ifx\csname l@#1\endcsname\relax
\typeout{** WARNING: IEEEtran.bst: No hyphenation pattern has been}%
\typeout{** loaded for the language `#1'. Using the pattern for}%
\typeout{** the default language instead.}%
\else
\language=\csname l@#1\endcsname
\fi
#2}}
\providecommand{\BIBdecl}{\relax}
\BIBdecl

\bibitem{Zheng2003}
L.~Zheng and D.~N.~C. Tse, ``Diversity and multiplexing: A fundamental tradeoff
  in multiple-antenna channels,'' \emph{IEEE Trans. Info. Theory}, vol.~49,
  no.~5, pp. 1073--1096, May 2003.

\bibitem{Nosratinia2004}
A.~Nosratinia, T.~E. Hunter, and A.~Hedayat, ``Cooperative communication in
  wireless networks,'' \emph{IEEE Commun. Mag.}, vol.~42, no.~10, pp. 74--80,
  Oct. 2004.

\bibitem{Laneman2004}
J.~N. Laneman, D.~N.~C. Tse, and G.~W. Wornell, ``Cooperative diversity in
  wireless networks: efficient protocols and outage behavior,'' \emph{IEEE
  Trans. Info. Theory}, vol.~50, no.~12, pp. 3062--3080, Dec. 2004.

\bibitem{Laneman2003}
J.~N. Laneman and G.~W. Wornell, ``Distributed space{-}time coded protocols for
  exploiting cooperative diversity in wireless networks,'' \emph{IEEE Trans.
  Info. Theory}, vol.~49, no.~10, pp. 2415--2425, Oct. 2003.

\bibitem{Sendonaris2003}
A.~Sendonaris, E.~Erkip, and B.~Aazhang, ``User cooperation diversity part {I}
  and part {II},'' \emph{IEEE Trans. Commun.}, vol.~51, no.~11, pp. 1927--1948,
  Nov. 2003.

\bibitem{Kramer2005}
G.~Kramer, M.~Gastpar, and P.~Gupta, ``Cooperative strategies and capacity
  theorems for relay networks,'' \emph{IEEE Trans. Info. Theory}, vol.~51,
  no.~9, pp. 3037--3063, Sept. 2005.

\bibitem{SSL2007}
A.~K. Sadek, W.~Su, and K.~J.~R. Liu, ``Multinode cooperative communications in
  wireless networks,'' \emph{IEEE Trans. Signal Processing}, vol.~55, no.~1,
  pp. 341--355, Jan. 2007.

\bibitem{shan2009}
H.~Shan, W.~Zhuang, and Z.~Wang, ``Distributed cooperative mac for multihop
  wireless networks,'' \emph{IEEE Commun. Mag.}, vol.~47, no.~2, pp. 126--133,
  Feb. 2009.

\bibitem{Ibrahim2008}
A.~S. Ibrahim, A.~K. Sadek, W.~Su, and K.~J.~R. Liu, ``Cooperative
  communications with relay selection: when to cooperate and whom to cooperate
  with?'' \emph{IEEE Trans. Wirel. Commun.}, vol.~7, no.~7, pp. 2814--2827,
  July 2008.

\bibitem{Zhao2005}
B.~Zhao and M.~C. Valenti, ``Practical relay networks: A generalization of
  hybrid-\mbox{ARQ},'' \emph{IEEE J. Sel. Areas Commun.}, vol.~23, no.~1, pp.
  7--18, Jan. 2005.

\bibitem{Dianati2006}
M.~Dianati, X.~Ling, K.~Naik, and X.~Shen, ``A node-cooperative {ARQ} scheme
  for wireless ad hoc networks,'' \emph{IEEE Trans. Vehic. Tech.}, vol.~55,
  no.~3, pp. 1032--1044, May 2006.

\bibitem{Ng2007}
T.~C.-Y. Ng and W.~Yu, ``Joint optimization of relay strategies and resource
  allocations in cooperative cellular networks,'' \emph{IEEE J. Sel. Areas
  Commun.}, vol.~25, no.~2, pp. 328--339, Feb. 2007.

\bibitem{YK2008}
Z.~Yi and I.-M. Kim, ``Diversity order analysis of the decode-and-forward
  cooperative networks with relay selection,'' \emph{IEEE Trans. Wirel.
  Commun.}, vol.~7, no.~5, pp. 1792--1799, May 2008.

\bibitem{syue09}
C.-L. Wang and S.-J. Syue, ``A geographic-based approach to relay selection for
  wireless ad hoc relay networks,'' in \emph{Proc. 2009 IEEE Vehic. Tech. Conf.
  (VTC 2009-Spring)}, Barcelona, Spain, April 2009, pp. 1--5.

\bibitem{Liu2006}
P.~Liu, Z.~Tao, Z.~Lin, E.~Erkip, and S.~Panwar, ``Cooperative wireless
  communications: A cross-layer approach,'' \emph{IEEE Trans. Wirel. Commun.},
  vol.~13, no.~4, pp. 84--92, Aug. 2006.

\bibitem{Beres2008}
E.~Beres and R.~Adve, ``Selection cooperation in multi-source cooperative
  networks,'' \emph{IEEE Trans. Wirel. Commun.}, vol.~7, no.~1, pp. 118--127,
  Jan. 2008.

\bibitem{Bletsas2005}
A.~Bletsas, A.~Lippman, and D.~P. Reed, ``A simple distributed method for relay
  selection in cooperative diversity wireless networks, based on reciprocity
  and channel measurements,'' in \emph{Proc. 2005 IEEE Vehic. Tech. Conf. (VTC
  2005-Spring)}, May 2005, pp. 1484--1488.

\bibitem{Gletsas2006}
A.~Gletsas, A.~Khisti, D.~P. Reed, and A.~Lippman, ``A simple cooperative
  diversity method based on network path selection,'' \emph{IEEE J. Sel. Areas
  Commun.}, vol.~24, no.~3, pp. 659--672, Mar. 2006.

\bibitem{Guo2008}
T.~Guo, R.~Carrasco, and W.~L. Woo, ``Performance of a cooperative relay-based
  auto-rate mac protocol for wireless ad hoc networks,'' in \emph{Proc. 2008
  IEEE Vehic. Tech. Conf. (VTC 2008-Spring)}, May 2008, pp. 11--15.

\bibitem{Liu2005}
P.~Liu, Z.~Tao, and S.~Panwar, ``A cooperative mac protocol for wireless local
  area networks,'' in \emph{Proc. 2005 IEEE Int. Conf. Commun. (ICC 2005)},
  Seoul, Korea, May 2005, pp. 2962--2968.

\bibitem{Zhu2006}
H.~Zhu and G.~Cao, ``\mbox{rDCF}: A relay-enabled medium access control
  protocol for wireless ad hoc networks,'' \emph{IEEE Trans. Mobile Comput.},
  vol.~5, no.~9, pp. 1201--1214, Sept. 2006.

\bibitem{CTBTMA08}
H.-S. Shan, W.~Wang, W.~Zhuang, and Z.~Wang, ``Cross-layer cooperative triple
  busy tone multiple access for wireless networks,'' in \emph{Proc. 2008 IEEE
  Global Commun. Conf. (GLOBECOM 2008)}, New Orleans, LA, USA, Dec. 2008, pp.
  1--5.

\bibitem{Gokturk2009}
M.~S. Gokturk and O.~Gurbuz, ``Cooperative mac protocol with distributed relay
  actuation,'' in \emph{Proc. 2009 IEEE Wireless Commun. and Networking Conf.
  (WCNC 2009)}, Budapest, Hungary, Apr. 2009, pp. 1--6.

\bibitem{Yu2010}
C.-H. Yu, O.~Tirkkonen, and J.~Hamalainen, ``Opportunistic relay selection with
  cooperative macro diversity,'' \emph{EURASIP J. Wirel. Commun. Networking},
  vol. 2010, pp. 1--14, 2010.

\bibitem{Wei2010}
Y.~Wei, F.~R. Yu, and M.~Song, ``Distributed optimal relay selection in
  wireless cooperative networks with finite-state markov channels,'' \emph{IEEE
  Trans. Vehic. Tech.}, vol.~59, no.~5, pp. 2149--2158, June 2010.

\bibitem{Larsson2001}
P.~Larsson, ``Selection diversity forwarding in a multihop packet radio network
  with fading channel and capture,'' \emph{SIGMOBILE Mob. Comput. Commun.
  Rev.}, vol.~5, no.~4, pp. 47--54, 2001.

\bibitem{Souryal2005}
M.~R. Souryal, B.~R. Vojcic, and R.~L. Pickholtz, ``Information efficiency of
  multihop packet radio networks with channel-adaptive routing,'' \emph{IEEE J.
  Sel. Areas Commun.}, vol.~23, no.~1, pp. 40--50, Jan. 2005.

\bibitem{SRM2009}
J.~A. Sanchez, P.~M. Ruiz, and R.~Marin-Perez, ``Beacon-less geographic routing
  made practical: challenges, design guidelines, and protocols,'' \emph{IEEE
  Commun. Magazine}, vol.~47, no.~8, pp. 85--91, Aug. 2009.

\bibitem{Heissen2003}
M.~Heissenb\"{u}ttel, T.~Braun, T.~Bernoulli, and M.~W\"{a}lchli, ``{BLR}:
  Beacon-less routing algorithm for mobile ad-hoc networks,'' \emph{Computer
  Commun.}, vol.~27, no.~11, pp. 1076--1086, 2003.

\bibitem{Fubler2003}
H.~Fubler, J.~Widmer, M.~Kasemann, M.~Mauve, and H.~Hartenstein,
  ``Contention-based forwarding for mobile ad hoc networks,'' \emph{Ad Hoc
  Networks}, vol.~1, no.~4, pp. 351--369, Nov 2003.

\bibitem{ZR2003}
M.~Zorzi and R.~R. Rao, ``Geographic random forwarding \mbox{(GeRaF)} for ad
  hoc and sensor networks: multihop performance,'' \emph{IEEE Trans. Mobile
  Comput.}, vol.~2, no.~4, pp. 337--348, Oct.-Dec. 2003.

\bibitem{Blum2003}
B.~Blum, T.~He, S.~Son, and J.~Stankovic, ``{IGF}: A state-free robust
  communication protocol for wireless sensor networks,'' in \emph{Tech. Rep.
  CS-2003-11}, Univ. of Virginia, Charlottesville, VA, 2003.

\bibitem{Casari2005}
P.~Casari, A.~Marcucci, M.~Nati, C.~Petrioli, and M.~Zorzi, ``A detailed
  simulation study of geographic random forwarding {GeRaF)} in wireless sensor
  networks,'' in \emph{Proc. 2005 IEEE Military Commun. Conf. (MILCOM 2005)},
  Oct. 2005, pp. 59--68.

\bibitem{Sanchez2007}
J.~A. Sanchez, R.~Marin-Perez, and P.~M. Ruiz, ``{BOSS}: Beacon-less on demand
  strategy for geographic routing in wireless sensor networks,'' in \emph{Proc.
  4th IEEE Int. Conf. on Mobile Ad-hoc and Sensor Systems (MASS 2007)}, Pisa,
  Italy, Oct. 2007, pp. 1--10.

\bibitem{AGSGAW2010}
T.~Aguilar, M.~C. Ghedira, S.-J. Syue, V.~Gauthier, H.~Afifi, and C.-L. Wang,
  ``A cross-layer design based on geographic information for cooperative
  wireless networks,'' in \emph{Proc. 2010 IEEE Vehic. Tech. Conf. (VTC
  2010-Spring)}, Taipei, Taiwan, May 2010.

\bibitem{HZF2004}
P.~Herhold, E.~Zimmermann, and G.~Fettweis, ``A simple cooperative extension to
  wireless relaying,'' in \emph{Proc. Int. Zurich Seminar Commun.}, Feb. 2004,
  pp. 36--39.

\bibitem{Su08}
W.~Su, A.~K. Sadek, and K.~J. Ray~Liu, ``Cooperative communication protocols in
  wireless networks: Performance analysis and optimum power allocation,''
  \emph{Wirel. Pers. Commun.}, vol.~44, no.~2, pp. 181--217, 2008.

\bibitem{Kalosha2008}
H.~Kalosha, A.~Nayak, S.~Ruhrup, and I.~Stojmenovic, ``Select-and-protest-based
  beaconless georouting with guaranteed delivery in wireless sensor networks,''
  in \emph{Proc. INFOCOM 2008}, Phoenix, AZ, USA, Apr. 2008, pp. 346--350.

\end{thebibliography}
\bibliographystyle{IEEEtran}
\vfill
\begin{IEEEbiography}[{\includegraphics[width=1in,height=1.25in,clip,keepaspectratio]{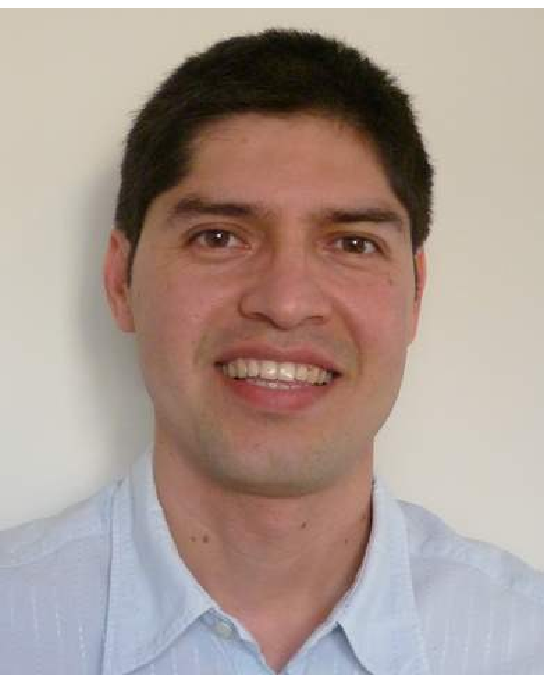}}]{Teck Aguilar}
was born in Chiapas, Mexico in 1974. He received the B.S. degree from the University of Mexico (UNAM), in 1998,
and the M.S. degree from the University of Paris 11, in 2005, both in computing engineering. Then, he joined Telecom
SudParis and the CNRS SAMOVAR (UMR 5157) lab, Evry, France, where his research interests pointed to cross-layer design in wireless sensor and ad hoc networks and cooperative communications. In 2010, he received his joint PhD degree in computing engineering from the University of Paris 6 and Telecom SudParis. Currently, he is a network consultant at a telecommunication company.
\end{IEEEbiography}

\begin{IEEEbiography}[{\includegraphics[width=1in,height=1.25in,clip,keepaspectratio]{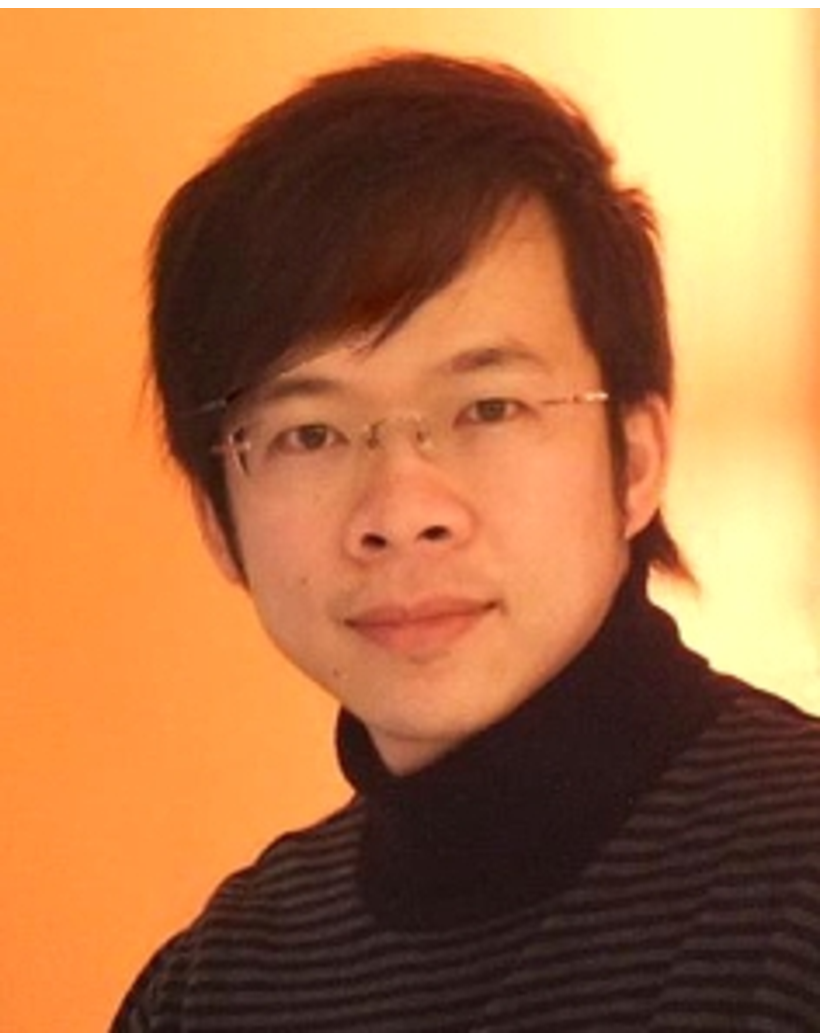}}]{Syue-Ju Syue}
(S\mbox{'}09) was born in Taipei, Taiwan, R.O.C., in 1981. He received the B.S. and M.S. degrees in electrical engineering from National Dong Hwa University, Taiwan, in 2003 and 2005, respectively. He is currently working toward the Ph.D. degree at National Tsing Hua University, Hsinchu, Taiwan. He was a visiting student at the Department of Wireless Networks and Multimedia Services, Telecom SudParis, Evry, France, during the spring semester of 2009. His research interests include cooperative communications and networking with special emphasis on relay selection, cooperative routing, and cross-layer design. Mr. Syue was awarded Honorary Member by Phi Tau Phi Scholastic Honor Society, R.O.C., in 2005.
\end{IEEEbiography}

\begin{IEEEbiography}[{\includegraphics[width=1in,height=1.25in,clip,keepaspectratio]{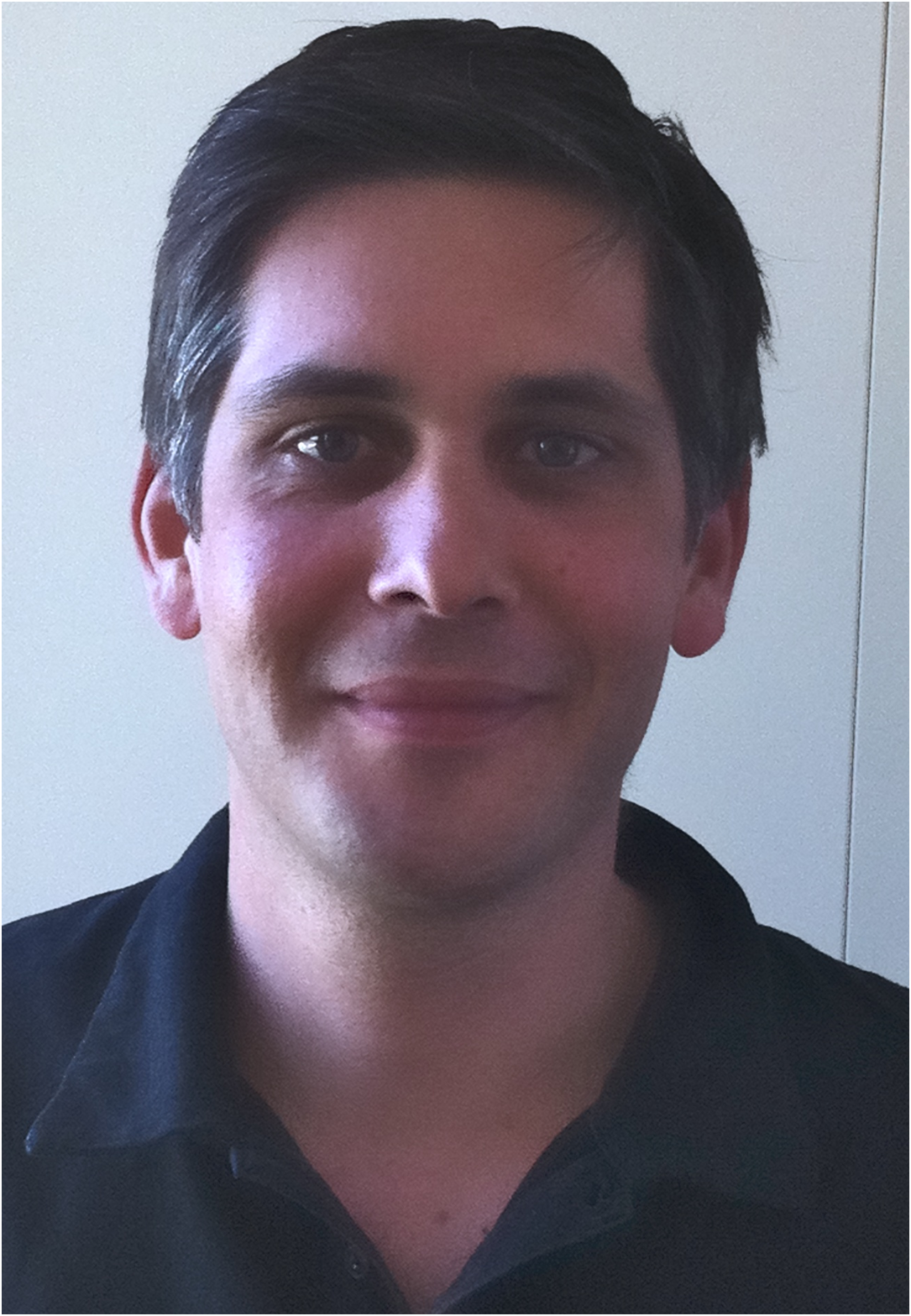}}]{Vincent Gauthier}
(S\mbox{'}03--M\mbox{'}06) was born in Paris in 1978. He received the B.S. in electrical engineering from University de Bretagne Occidentale in 2002 and M.S. degrees the Ph.D. degree in electrical engineering and computer networks  from University of Paris 6 in 2003 and 2006, respectively. He was a Guest Researcher at National Institute of Standards and Technology, MA, United States between 2006 and 2008. He joined the faculty of Telecom SudParis, and the lab CNRS SAMOVAR (UMR 5157), Evry, France, in 2008, where he is currently an Associate Professor of the Department of Wireless Networks and Multimedia Services. His current research interests are primarily on wireless networks, sensor networks, ad-hoc networks, cross-layer design, self-organization in wireless networks, and cooperative communications. His other research interests include mobility modeling, performance analysis, and queuing theory.
\end{IEEEbiography}

\begin{IEEEbiography}[{\includegraphics[width=1in,height=1.25in,clip,keepaspectratio]{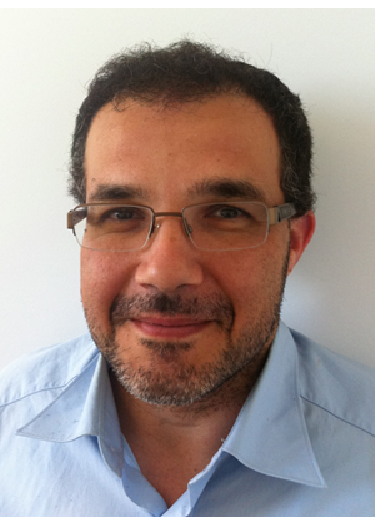}}]{Hossam Afifi}
is a professor at Telecom SudParis and the lab CNRS SAMOVAR (UMR 5157), where he works on mobile and security protocols in wireless communications. His current interests cover vehicular and user centric wireless systems. He obtained the Ph.D. from Inria Sophia Antipolis in the field of computer science in 1992. He visited Washington University, St. Louis as a Post Doc where he worked on IP Switching techniques. Hossam was appointed as an assistant professor at ENST Bretagne, France in the field of high speed networking. After his tenure and a sabbatical in Nokia Research Labs, Mountain View USA, he took the current position at Telecom SudParis in 2000.
\end{IEEEbiography}
\vfill
\begin{IEEEbiography}[{\includegraphics[width=1in,height=1.25in,clip,keepaspectratio]{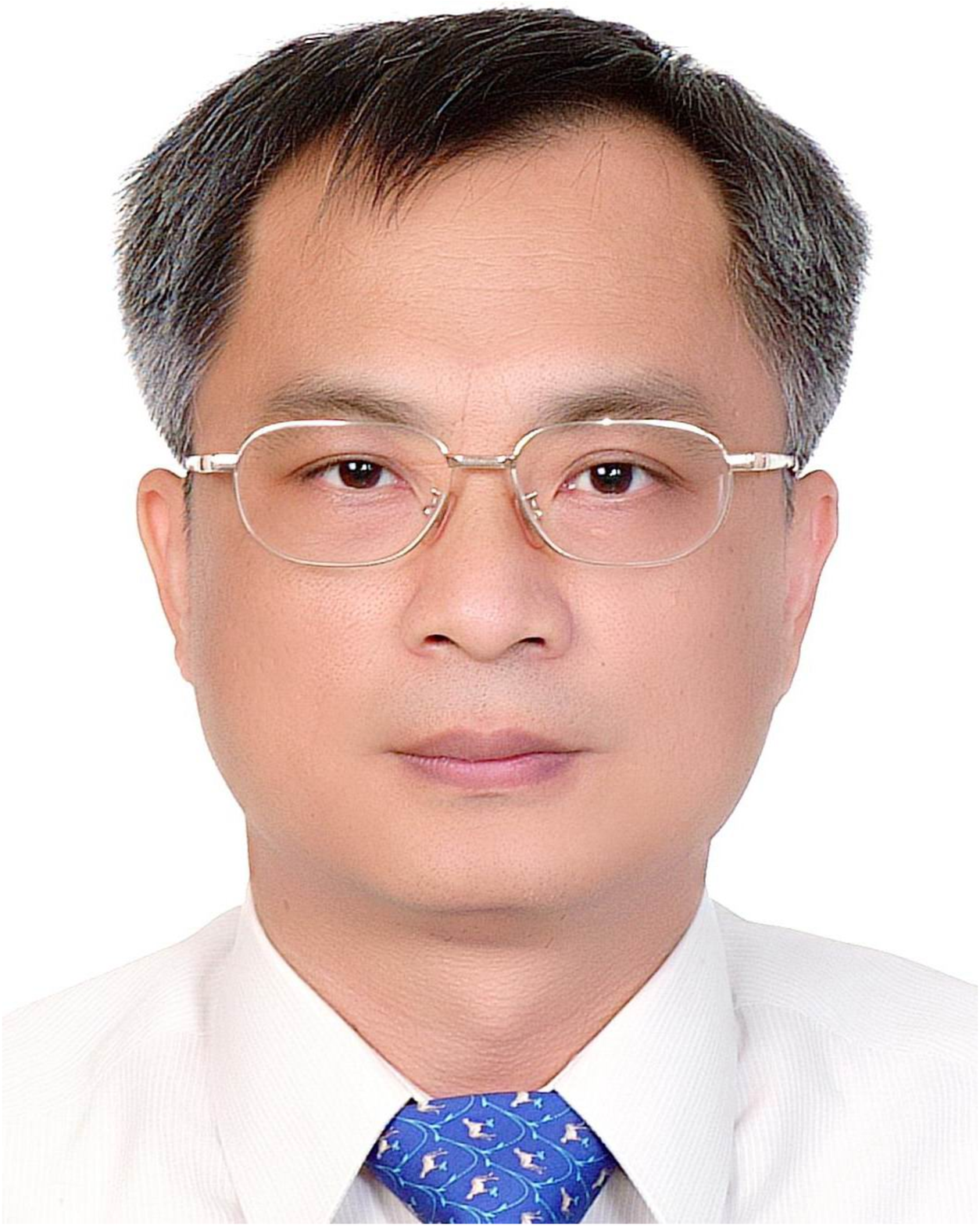}}]{Chin-Liang Wang}
(S\mbox{'}85--M\mbox{'}87--SM\mbox{'}04) was born in Tainan, Taiwan, R.O.C., in 1959. He received the B.S. degree in electronics engineering from National Chiao Tung University (NCTU), Hsinchu, Taiwan, in 1982, the M.S. degree in electrical engineering from National Taiwan University, Taipei, Taiwan, in 1984, and the Ph.D. degree in electronics engineering from NCTU in 1987.

He joined the faculty of National Tsing Hua University (NTHU), Hsinchu, Taiwan, in 1987, where he is currently a Professor of the Department of Electrical Engineering and the Institute of Communications Engineering. During the academic year 1996-1997, he was on leave at the Information Systems Laboratory, Department of Electrical Engineering, Stanford University, Stanford, CA, as a Visiting Scholar. He served as the Director of the Institute of Communications Engineering from 1999 to 2002 and the Director of the University’s Computer \& Communications Center from 2002 to 2006. He was the Chair of the Wireless Networks Group of the National Science \& Technology Program for Telecommunications from 2004 to 2008, and has been the Chair of the Access Technology Group of the Networked Communications Program since 2009. He is also serving as the Director of the Communications Engineering Program, National Science Council, R.O.C. His current research interests are primarily in baseband technologies for wireless communications and wireless sensor networks.

Dr. Wang was a recipient of the Distinguished Teaching Award granted by the Ministry of Education, R.O.C., in 1992 and the Distinguished Electrcial Engineering Professor Award granted by the Chinese Institute of Electrical Engineering in 2010. He received the Acer Dragon Thesis Award in 1987 and the Acer Dragon Thesis Advisor Awards in 1995 and 1996. In the academic years 1993-1994 and 1994-1995, he received the Outstanding Research Awards from the National Science Council, R.O.C. He received the HDTV Academic Achievement Award from the Digital Video Industry Development Program Office, Ministry of Economic Affairs, R.O.C., in 1996. He was also the advisor on several technical works that won various awards in Taiwan, including the Outstanding Award of the 1993 Texas Instruments DSP Product Design Challenge in Taiwan, the Outstanding Award of the 1994 Contest on Design and Implementation of Microprocessor Application Systems sponsored by the Ministry of Education, R.O.C., the Outstanding Award of the 1995 Student Paper Contest sponsored by the Chinese Institute of Engineers, and the 1995 and 2000 Master Thesis Awards of the Chinese Institute of Electrical Engineering. He served as an Associate Editor for the IEEE Transactions on Signal Processing from 1998 to 2000 and has been an Editor for Equalization for the IEEE Transactions on Communications since 1998. He was also one of the Guest Editors for the Special Issue on Model Order Selection in Signal Processing Systems of the IEEE Journal of Selected Topics in Signal Processing.
\end{IEEEbiography}
\vfill
\end{document}